\documentclass[twocolumn,a4paper,nofootinbib,superscriptaddress]{revtex4}

\usepackage{amsmath,amssymb,graphicx,epstopdf,enumerate,amsfonts,booktabs,color,hyperref}
\usepackage{float,placeins,lscape}
\usepackage{relsize}
\usepackage{soul}
\usepackage[margin=1.2cm]{geometry}
\usepackage[caption=false]{subfig}

\begin{document}

\title{QCD Phase Diagram by Holography}
\author{Yi Yang}
\email{yiyang@mail.nctu.edu.tw}
\affiliation{Department of Electrophysics, National Chiao Tung University, Hsinchu, ROC}
\affiliation{Physics Division, National Center for Theoretical Sciences, Hsinchu, ROC}
\author{Pei-Hung Yuan}
\email{phy.pro.phy@gmail.com}
\affiliation{School of physics, University of Chinese Academy of Sciences, Beijing
100049, China}
\affiliation{Kavli Institute for Theoretical Sciences, University of Chinese Academy of
Sciences, Beijing 100049, China}

\begin{abstract}
We explore QCD phase diagram by constructing a holographic QCD model using the Einstein-Maxwell-Scalar system. The chiral transition is investigated by adding a probe scalar and confinement transition is studied by adding a probe string into the system. By interpreting the black hole phase transition in the bulk spacetime as the quarkyonic transition in the dual QCD theory and introducing the bypass mechanism for deconfinement transition, we give an explanation why chiral symmetry breaking and deconfinement transition lines coincide with each other despite their different physical origins.
\end{abstract}

\maketitle

\setcounter{equation}{0}

\noindent \textit{Introduction} It is widely believed that Quantum chromodynamics (QCD) is in the confinement and chiral symmetry breaking ($\chi$SB) phase at low temperature and small chemical potential region, while in the deconfinement and chiral symmetry restored ($\chi$S) phase at high temperature and large chemical potential territory. It is then conjectured that there is a phase transition between these two phases \cite{Cabibbo}. However, deconfinement and $\chi$SB have quite different physical origins. In principle, the deconfinement transition is defined in the quench limit $(m_{q}\rightarrow \infty)$ with the Polyakov loop as its order parameter, while the $\chi$SB is defined in the chiral limit $(m_{q}\rightarrow 0)$ with the quark condensation as its order parameter. In practice, it is difficult to determine the QCD phase diagram due to the strong interaction near the phase transition region where the conventional perturbation method does not work.

For a few decades, lattice QCD simulation (LQCD) is the only reliable method to attack this problem. Around zero chemical potential $\mu \sim 0$, LQCD showed that the transition temperatures for the deconfinement and $\chi$SB are almost the same \cite{0609068}, and the phase transition reduces to a crossover for the physical quark mass \cite{1111.4953} that implies a critical end point (CEP) at certain chemical potential $\mu$. However, LQCD suffers the sign problem for finite chemical potentials.

\begin{figure}[t]
\subfloat[QCD phase diagram in Large $N_c$ limit]{\includegraphics[width=.8\linewidth]{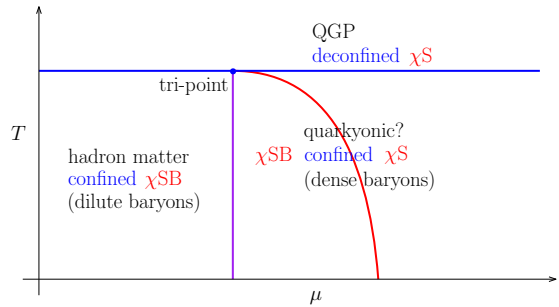}}\\
\subfloat[Conjectured QCD
phase diagram for finite $N_{c}$]{\includegraphics[width=.8\linewidth]{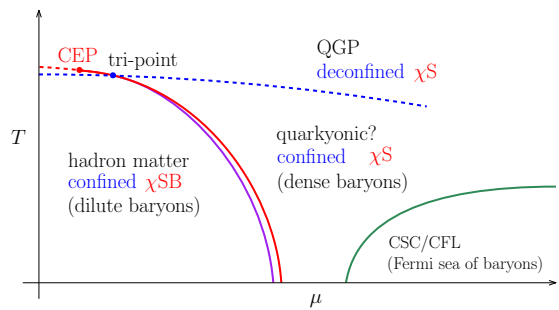}}
\caption{The blue, purple and red lines label the deconfinement, $\chi$SB and quarkyonic transition, respectively.}
\label{Nphase}
\end{figure}

Moreover, large $N_{c}$ QCD provides an expressive picture for the QCD phase diagram as shown in FIG.\ref{Nphase}(a). In the large $N_{c}$ limit, the transition temperature of the deconfinement phase transition $T_{d}$ (blue line) is independent of the chemical potential since quarks do not affect the gluons in this limit. Above $T_{d}$, QCD is in the quark gluon plasma (QGP) phase with deconfinement and $\chi$S. Below $T_{d}$, QCD is in the hadronic phase with confinement and $\chi$SB at small $\mu$, and in the quarkyonic phase at large $\mu$. There is a quarkyonic phase transition (purple line) between the two phases below $T_{d}$ with the baryon density as its order parameter \cite{0706.2191}. The hadronic, quarkyonic and QGP phases encounter at a tri-point $(\mu_{t}, T_t)$. Furthermore, it has been argued that the $\chi$SB transition (red line) is in the quarkyonic phase \cite{0706.2191}.

However, for finite $N_{c}$, it was argued that the deconfinement transition would reduce to a crossover \cite{1801.09215} and decrease with the chemical potential \cite{0701022} implying the shrink of the quarkyonic region \cite{0803.3318,0805.1509}. In addition, it was shown that the quarkyonic transition is close to the $\chi$SB with the CEP of $\chi$SB at $(\mu_{c}, T_c)$ being not far from the tri-point \cite{0911.4806}. LQCD also showed that the quarkyonic transition is in the deconfined phase for small $\mu$ \cite{0905.2949}. Moreover, at low $T$ and very large $\mu$, QCD is conjectured to be in the color superconductor (CSC) or color-flavor locking (CFL) phase \footnote{We will not consider CSC/CFL phase in this work.}. Combining the above evidences, the conjectured QCD phase diagram for finite $N_{c}$ is plotted in FIG.\ref{Nphase}(b). Nevertheless, whether or not the quarkyonic phase exists for finite $N_{c}$ is still an open question. Many recent studies suggested that the deconfinement and $\chi$SB coincide for $\mu \geq \mu_{t}$ implying the vanishing of the quarkyonic region \cite{1104.0873,1602.06699,1705.00718}, while others claimed that there is no direct relation between the deconfinement and $\chi$SB \cite{1405.1289,1502.07706,1709.05981}.

On the other hand, holographic correspondence \cite{9711200} offers an ideal framework to explore QCD in the strongly coupled region by studying its dual gravitational theory, i.e., holographic QCD (hQCD). Several hQCD models have been constructed to study the phase transitions in QCD \cite{1511.02721,1512.06493,1610.09814,1810.07019,1908.02000,1910.02383}. However, different conclusions have been obtained from various models.

In this work, we use an extensively studied hQCD model based on the Einstein-Maxwell-Scalar (EMS) system to investigate QCD phase structure. A probe scalar and open strings are added into the system to realize the $\chi$SB and deconfinement phase transitions, respectively. Nevertheless, the QCD dual of the black hole phase transition in the bulk spacetime is still not clear. In this work, we propose to interpret the black hole phase transition as the quarkyonic transition in QCD and show that it coincides with the $\chi$SB. Furthermore, we introduce the bypass mechanism in the holographic viewpoint and show that the quarkyonic transition also coincides with the deconfinement transition. We thus conclude that the dynamically stable quarkyonic state does not exist and explain why the deconfinement and $\chi$SB transitions coincide despite their different physical origins.\\

\noindent \textit{Holographic Model} We follow the hQCD model constructed in \cite{1301.0385,1406.1865,1506.05930,1703.09184,1705.07587,2004.01965,2009.05694}
by considering the 5-dimensional EMS system with probe matter fields $S=S_{b}+S_{\chi }+S_{NG}$. In string frame, labeled by the super-index s,
\begin{align}
&S_{b} \sim \int d^{5}x\sqrt{-g^s}e^{-2\phi^s}
	\left[{R^s-\frac{f^s(\phi^s) {F}^{2}}{4}}
			+4\vert\partial \phi^s\vert^2
			-V^s(\phi^s)  \right] ,
\label{eq_SB_sf} \\
&S_{\chi } \sim \int d^{5}x\sqrt{-g^s}e^{-\phi^s}Tr\{\nabla _{M}X^{\dagger
}\nabla ^{M}X+m_{\chi }^{2}X^{\dagger }X\},  \label{SX} \\
&S_{NG} \sim \int d^{2}\xi \sqrt{-G},\text{ }G_{ab}=g_{\mu \nu} ^s\partial
_{a}X^{\mu }\partial _{b}X^{\nu }.
\end{align}
where the background action $S_{b}$ includes a gravity field $g^s_{\mu \nu}$, a Maxwell field ${F}_{\mu \nu}=\partial_{\mu }{A}_{\nu }-\partial_{\nu }{A}_{\mu}$ and a neutral scalar field $\phi ^s$. The gauge kinetic function $f^s$ and the scalar potential $V^s$ are functions of the scalar field $\phi ^s$. $S_{\chi}$ is for a composite massive scalar $X$ describing the quark condensation $\langle \bar{\psi}\psi \rangle$ in the vacuum and $S_{NG}$ is the Nambu-Goto action describing open strings which is the test objects for the deconfinement transition. We treat all matter fields as probes by ignoring their backreaction.\\

\noindent \textit{Background Solutions} In practice, it is more convenience to study the background in Einstein frame by a Weyl transformation: $\phi^{s}=\sqrt{\frac{3}{8}}\phi$, $g^s_{\mu\nu}=g_{\mu\nu} {e}^{\sqrt{\frac{2}{3}}\phi}$, $f^{s}=f {e}^{\sqrt{\frac{2}{3}}\phi}$ and $V^{s}={e}^{-\sqrt{\frac{2}{3}}\phi}V$. The background action Eq.(\ref{eq_SB_sf}) becomes 
\begin{equation}
S_{b} \sim \int d^{5}x\sqrt{-g}\left[ {R-\frac{f(\phi)}{4}{F} ^{2}}- \frac{1}{2}\vert \partial \phi\vert^2 -V(\phi )\right]
\label{eq_SB_Ef}
\end{equation}
To study QCD at finite temperature and chemical potential, we consider an asymptotic AdS black hole ansatz,
\begin{align}
ds^{2}& =\frac{e^{2A\left( z\right) }}{z^{2}}\left[
-g(z)dt^{2}+d\vec{x}^{2}+\frac{dz^{2}}{g(z)}\right], \label{eq_metric} \\
\phi & =\phi \left( z\right),
~A_{\mu }=\left(A_{t}\left( z\right) ,\vec{0},0\right), \label{eq_ansatz}
\end{align}
where $z=0$ corresponds to the conformal boundary of the 5-dimensional bulk spacetime. The equations of motion then can be derived from the action Eq.(\ref{eq_SB_Ef}),
\begin{align}
A_{t}^{\prime \prime }+\left( \frac{f^{\prime }}{f}+A^{\prime }-\dfrac{1}{z}
\right) A_{t}^{\prime }& =0, \label{eom_At} \\
\dfrac{\phi ^{\prime 2}}{6}+A^{\prime \prime }-A^{\prime 2}+\dfrac{2}{z}
A^{\prime }& =0, \label{eom_A} \\
g^{\prime \prime }+3g^{\prime }\left( A^{\prime }-\dfrac{1}{z}\right) -\frac{
fz^{2}A_{t}^{\prime 2}}{e^{2A}}& =0. \label{eom_g}
\end{align}
Imposing the asymptotic AdS condition $A(0)=\phi (0)=0$, $g(0)=1$ at $z=0$ and the regular condition $A_{t}(z_{h})=g(z_{h})=0$ at $z=z_{h}$, the equations of motion Eqs.(\ref{eom_At}-\ref{eom_g}) can be solved analytically,
\begin{align}
\phi \left( z\right) & =\int_{0}^{z}dy\sqrt{-6\left( A^{\prime \prime}-A^{\prime 2}+\dfrac{2}{y}A^{\prime }\right) }, \label{phip-A} \\
A_{t}\left( z\right) & =\mu \left[ 1-\frac{I_{2}(z)}{I_{2}(z_{h})}\right]=\mu +\rho z^{2}+\cdots , \label{At-A} \\
g\left( {z}\right) & =1-\frac{I_{1}(z)}{I_{1}(z_{h})}+\dfrac{\mu^{2}\left\vert
\begin{array}{cc}
I_{1}(z_{h}) & \int_{0}^{z_{h}}I_{1}^{\prime }(y)I_{2}(y)dy \\
I_{1}(z) & \int_{0}^{z}I_{1}^{\prime }(y)I_{2}(y)dy%
\end{array}%
\right\vert }{I_{1}(z_{h})I_{2}^{2}(z_{h})},  \label{g-A} \\
I_{1}(z)& =\int_{0}^{z}\frac{y^{3}}{e^{3A}}dy\text{, }I_{2}(z)=\int_{0}^{z}%
\frac{y}{fe^{A}}dy,  \label{int-A}
\end{align}
where $\mu$ and $\rho$ represent chemical potential and baryon density by holographic correspondence. Eqs.(\ref{phip-A}-\ref{int-A}) are a family of solutions for different choices of the functions $A$ and $f$. To realize the linear Regge trajectories of vector meson spectrum and fit the chiral phase transition temperature at zero chemical potential $T_{\chi }^{0}\simeq 0.159$ from the recent lattice QCD simulations \cite{2002.02821}, we fix $f = e^{-cz^{2}-A-\sqrt{\frac{3}{8}}\phi }$ and $A=-a\ln (bz^{2}+1)$ with the parameters $a=4.037$, $b=0.0155$ and $c=0.227$ in the following \cite{1703.09184,1812.09676}.\\

\begin{figure}[tbp]
\subfloat[$T - z_h$]{\hspace{-.5cm} \includegraphics[width=.45\linewidth]{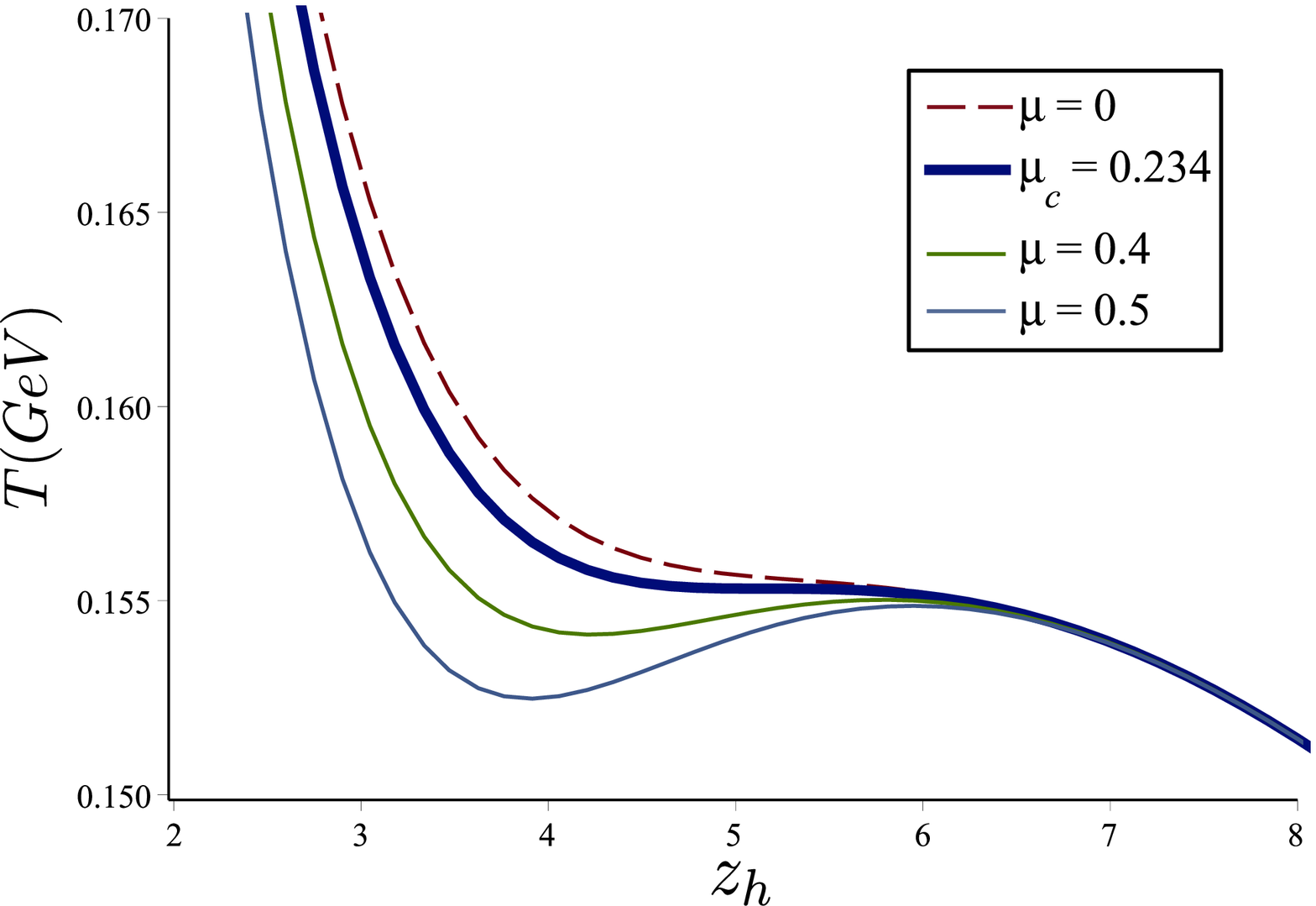}} %
\subfloat[$F - T$]{\includegraphics[width=.5\linewidth]{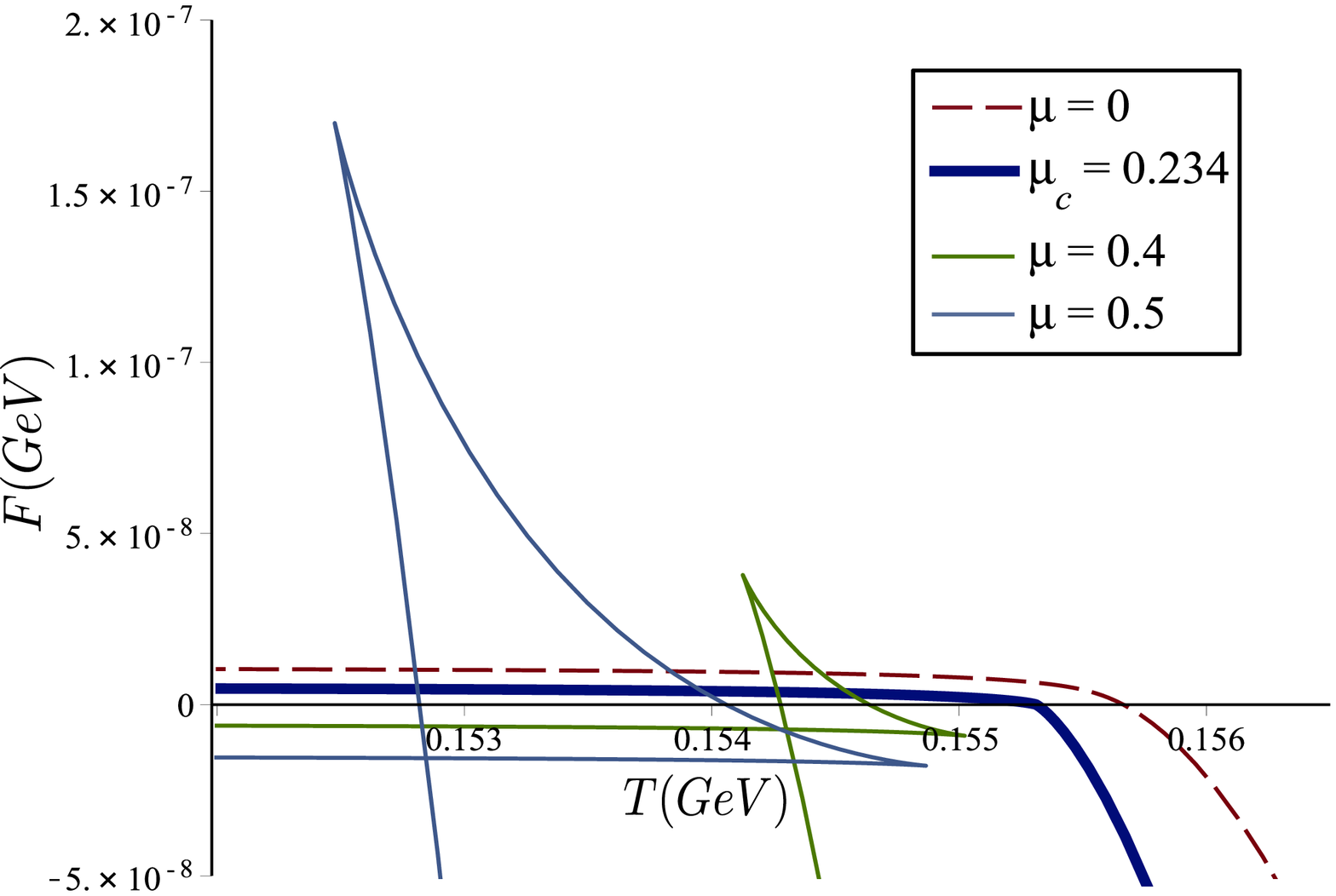}}\newline
\subfloat[$\rho - \mu$]{\includegraphics[width=.45\linewidth]{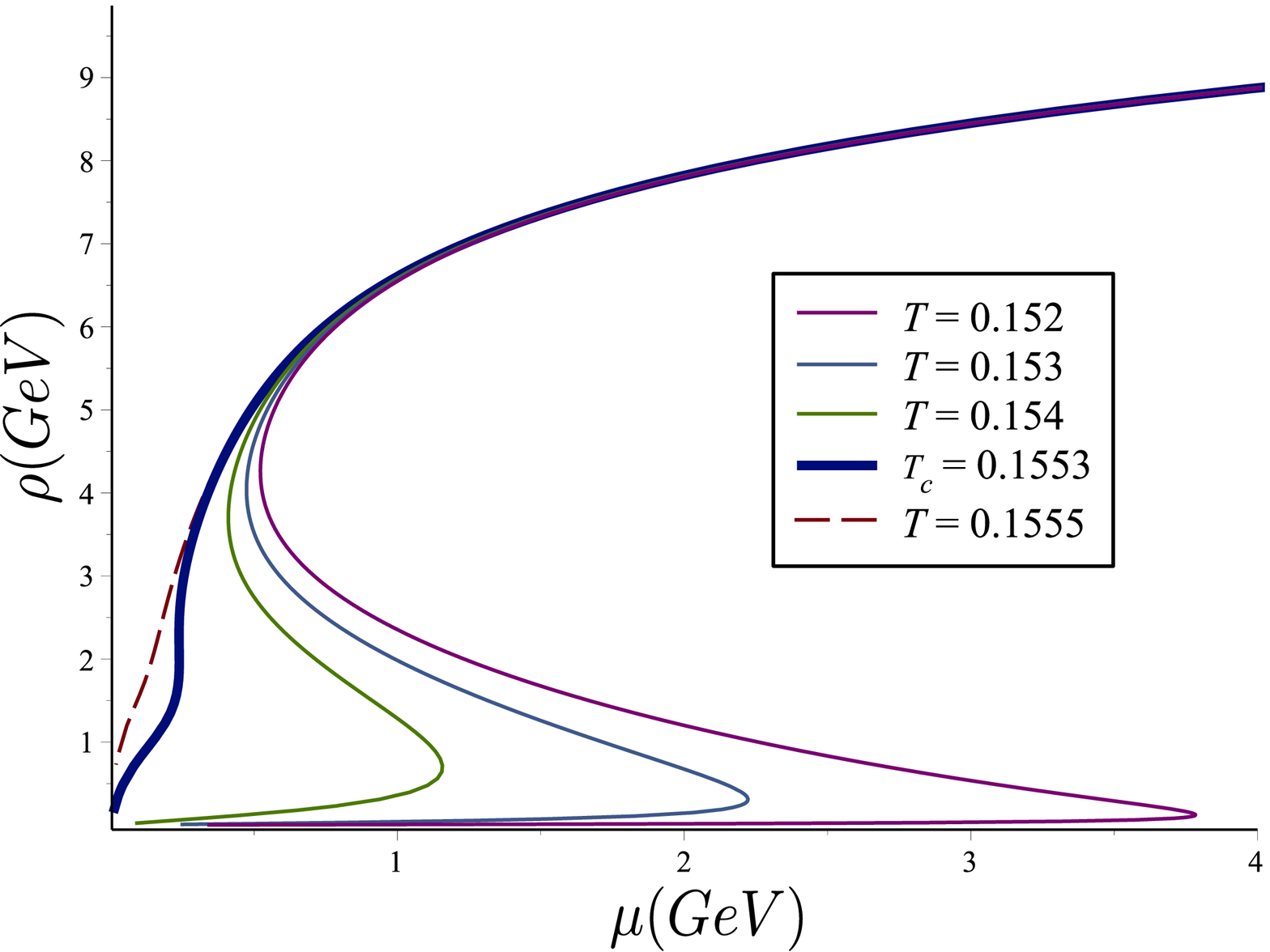}}
\subfloat[BH Phase
diagram]{\includegraphics[width=.5\linewidth]{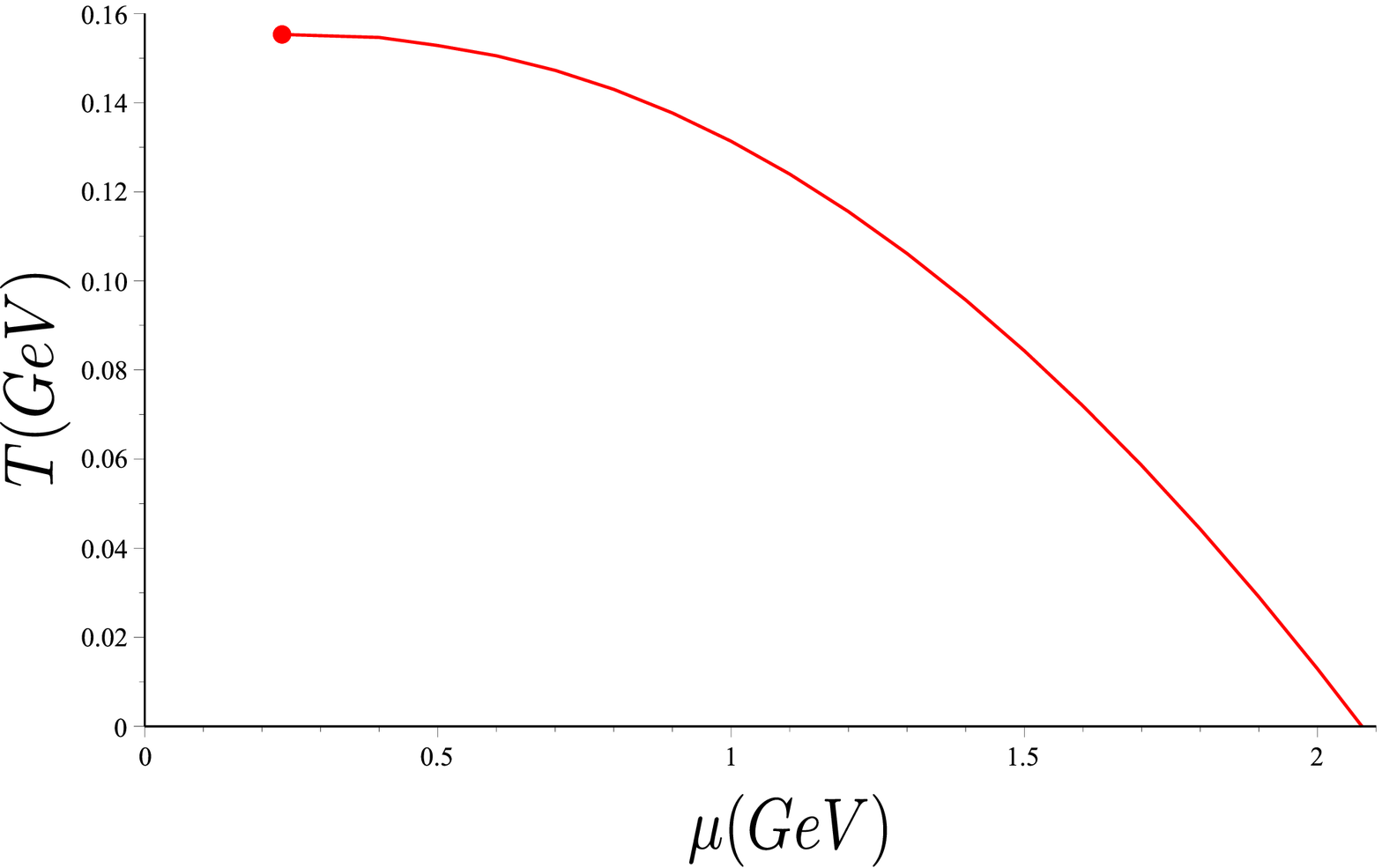}}
\caption{(a) Temperature vs. black hole horizon at different chemical potentials. (b) Free energy vs. temperature at different chemical potentials. (c) Baryon density vs. chemical potential at different temperature. (d) Black hole phase diagram in $T-\protect\mu$ plane with the critical endpoint at $(\protect\mu_{c},T_{c})\simeq (0.216,0.159)~GeV$.}
\label{fig_Tmu_BH}
\end{figure}

\noindent \textit{Black Hole Phase Transition} The black hole temperature $T=g^{\prime }\left( z_{h}\right) /4\pi$ is plotted in FIG.\ref{fig_Tmu_BH}(a). The multivalued behavior at $\mu >\mu_{c}$ implies a phase transition. The transition temperature can be obtained from the free energy $F=-\int sdT$ at each given chemical potential with the entropy density $s=e^{3A(z_{h})}/4z_{h}^{3}$. The free energy at different chemical potentials are plotted in FIG.\ref{fig_Tmu_BH}(b). At $\mu >\mu_{c}$, the swallow tail shape implies a first-order phase transition which reduces to a second-order one at the CEP $(\mu_{c},T_{c})$ and becomes a crossover for $\mu <\mu_{c}$. The black hole phase diagram is plotted in FIG.\ref{fig_Tmu_BH}(c).

How to interpret the black hole phase transition in the dual hQCD is still not clear. It seems natural to follow the idea of interpreting the Hawking-Page transition as the deconfinement transition in the dual field theory. However, since both phases permit a black hole in the current case, this interpretation cannot describe the confinement phase. We propose to interpret the black hole phase transition in the bulk spacetime as the quarkyonic transition in the dual hQCD. To justify our proposal, we calculate the baryon density $\rho =\mu /2I_{2}(z_{h})$, which is the order parameter of the quarkyonic transition, and plot it in FIG.\ref{fig_Tmu_BH}(d). The baryon density jumps from a tiny value to a huge amount during the black hole phase transition implying that the system transits to the quarkyonic phase.\\

\begin{figure}[tbp]
\begin{tabular}{cc}
\hspace{-.5cm} \begin{minipage}{0.3\textwidth}
\subfloat[]{\includegraphics[width=0.9\textwidth]{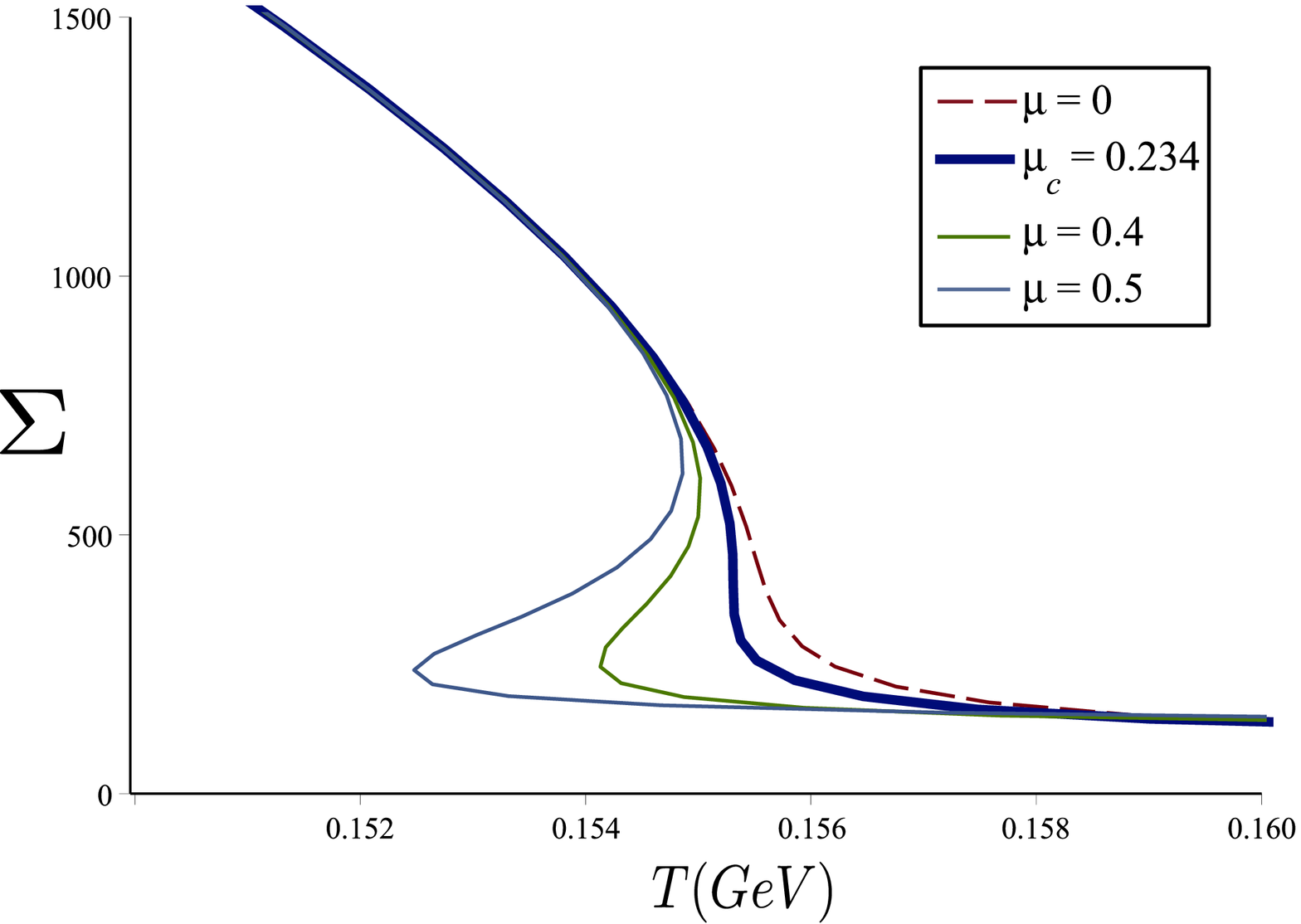}} \end{minipage}
& \hspace{-0.5cm} \begin{minipage}{0.2\textwidth} \subfloat[$\mu=0$]{
\begin{tabular}[b]{@{}c@{}} \includegraphics[width=.45\linewidth]{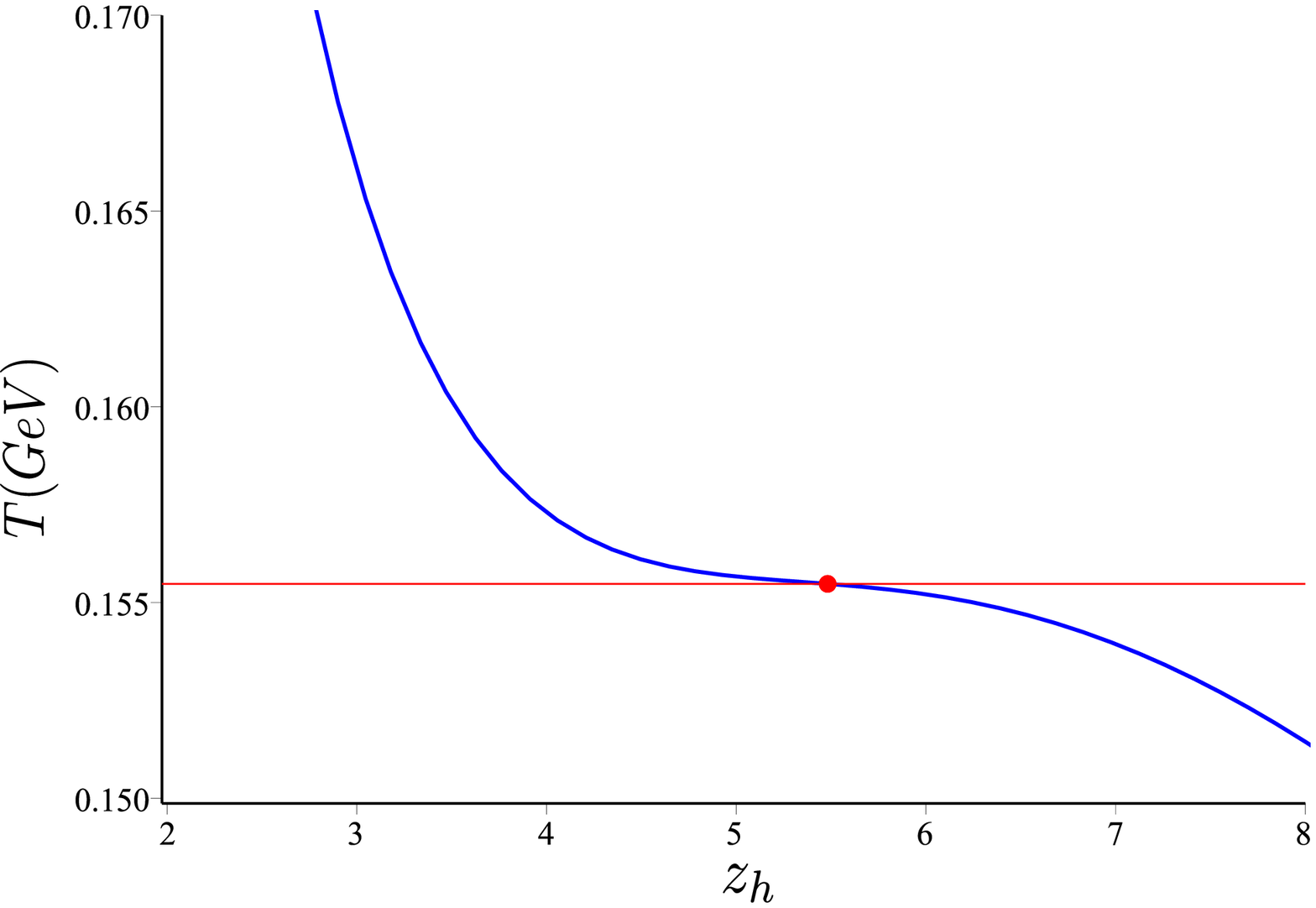}
\includegraphics[width=.45\linewidth]{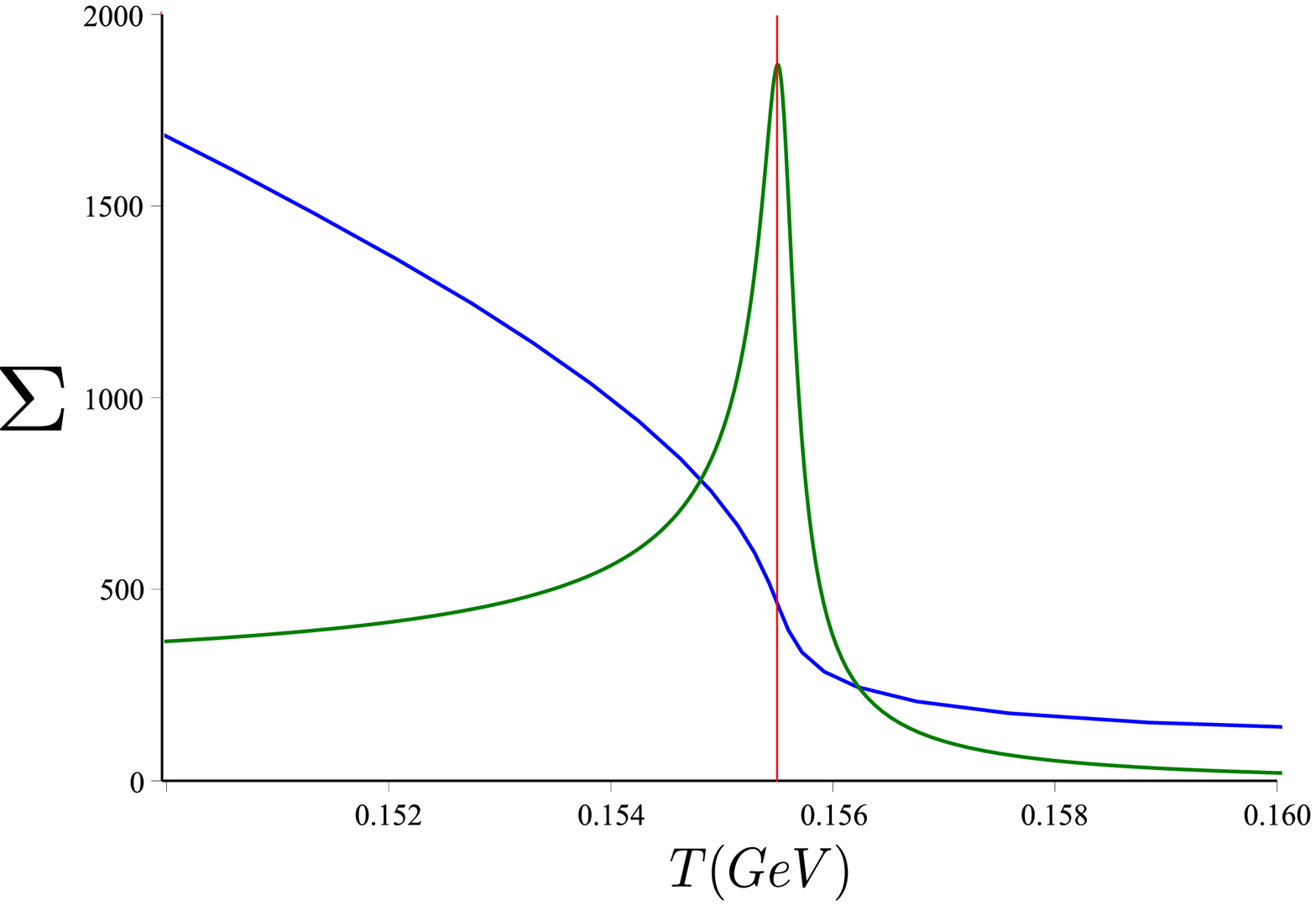} \end{tabular}}\\
\vspace{-0.3cm} \subfloat[$\mu=0.5$]{ \begin{tabular}[b]{@{}c@{}}
\includegraphics[width=.45\linewidth]{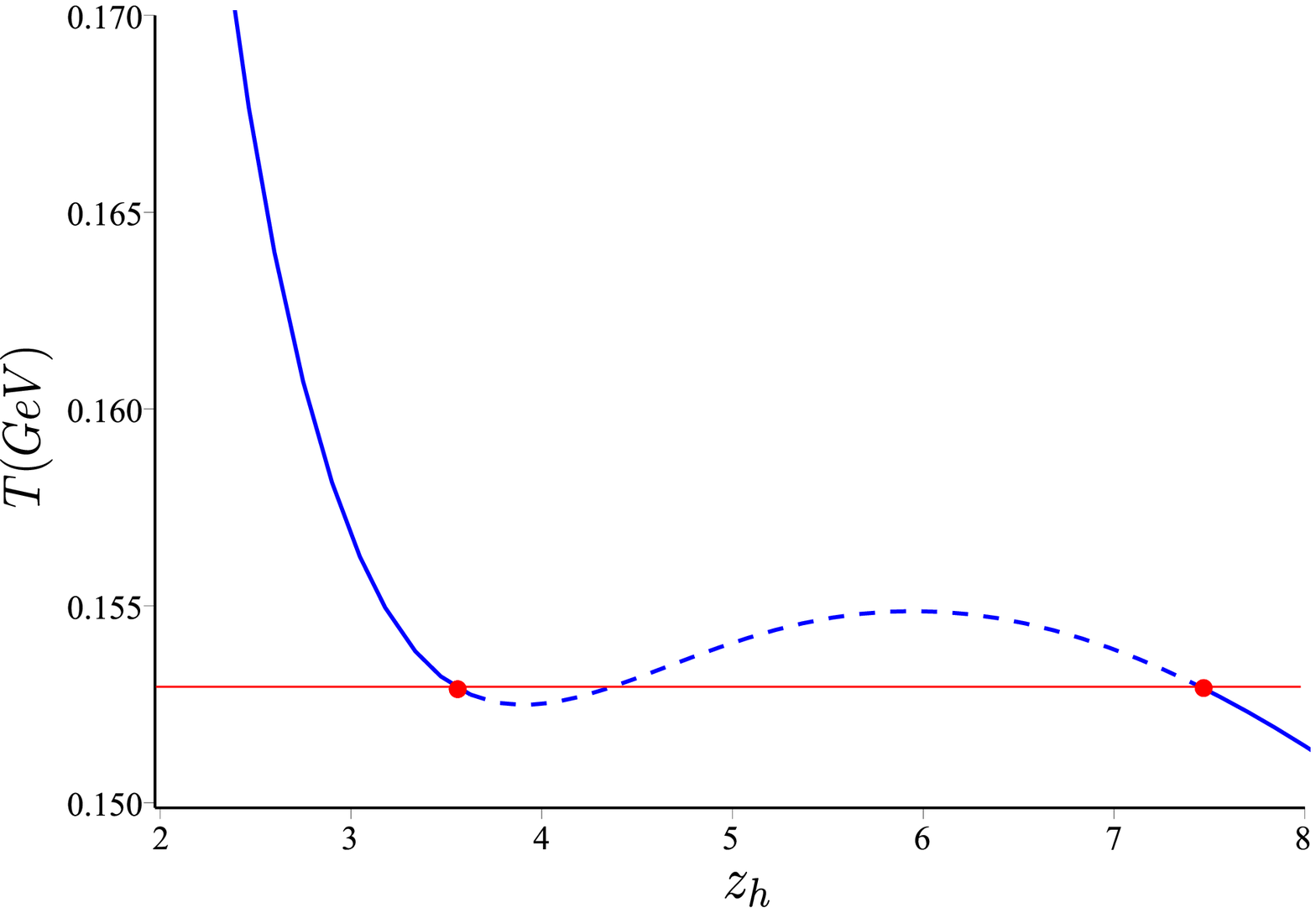}
\includegraphics[width=.45\linewidth]{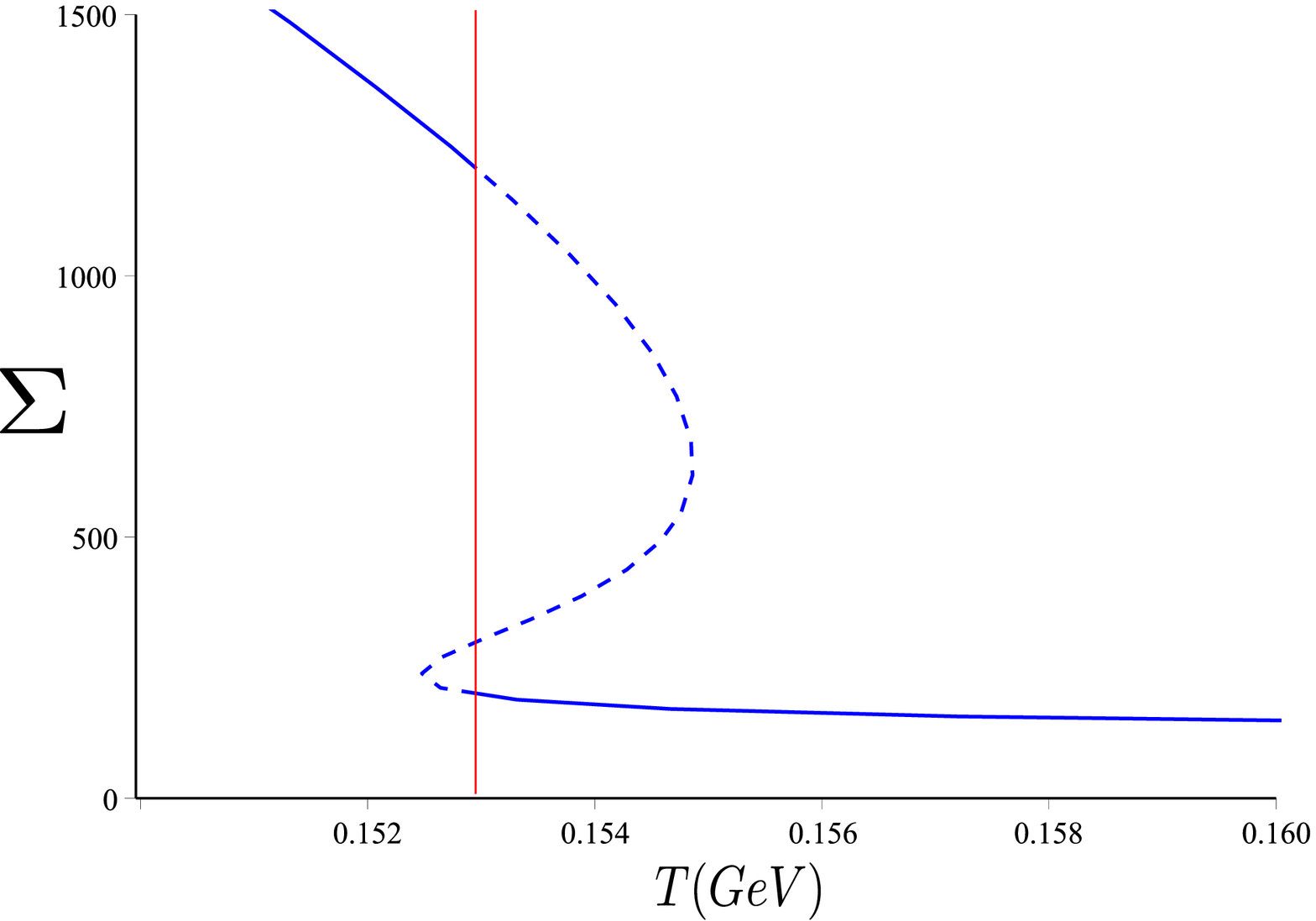} \end{tabular}}
\end{minipage}%
\end{tabular}%
\caption{(a) Quark condensation vs. temperature at different chemical potentials. (b, c) The $\protect\chi$SB transition temperatures at $\protect\mu =\{0,0.5\}~GeV$.}
\label{SigmaT}
\end{figure}

\noindent \textit{Chiral Symmetry Breaking} We consider $N_{f}=2$ in this work. The probe scalar field $X$ in the action Eq.(\ref{SX}) is a $2\times 2$ matrix and can be brought to the diagonal form $X_{ii}=\chi /2$. By varying the action Eq.(\ref{SX}), we obtain the equation of motion for the probe scalar field $\chi$,
\begin{equation}
\chi ^{\prime \prime }(z)+\frac{p(z)}{z}\chi ^{\prime }(z)-\frac{q(z)}{z^{2}} \chi (z)=0, \label{eom}
\end{equation}
where $p=z\left( 3A_{s}^{\prime }-\phi _{s}^{\prime }+g^{\prime }/g\right) -3 $ and $q=e^{2A_{s}}m_{\chi }^{2}/g$ are regular at $z=0$. And $A_s=A+\frac{\phi}{\sqrt{6}}$. Near the boundary at $z=0$, Eq.(\ref{eom}) can be solved by Frobenius method \cite{2009.05694},
\begin{align}
\chi _{B}(z)& =\alpha \chi _{1}(z)+\beta \chi _{2}(z)  \notag \\
& =\beta z+\beta C_{1}^{(2)}z^{2}+\alpha z^{3}+\beta Cz^{3}\ln z  \notag \\
& +(\alpha C_{1}^{(1)}+\beta C_{3}^{(2)})z^{4}+\beta CC_{1}^{(1)}z^{4}\ln
z+\cdots  \label{general chi}
\end{align}
where $\alpha$ and $\beta$ are coefficients of the linear combination and the coefficients $C,C_{i}^{(1)}$ and $C_{i}^{(2)}$ can be obtained order by order from Eq.(\ref{eom}). On the other hand, by holographic correspondence, the scalar field $\chi$ can be expanded at the boundary as 
\begin{equation}
\chi_{B}(z)=m_{q}\zeta z+\Sigma z^{3}/\zeta +\cdots \label{boundary chi}
\end{equation}
where $m_{q}\simeq 3MeV$ is the current quark mass, $\zeta =N_c/2\pi N_f^{1/2}$ with $N_c=3$ and $N_f=2$ is a constant, and $\Sigma=\langle \bar{\psi}\psi \rangle$ is the quark condensate.
Comparing Eq.(\ref{general chi}) and Eq.(\ref{boundary chi}), we have $\Sigma =\alpha \zeta$.

Near the horizon at $z=z_{h}$, the scalar field can be expanded as,
\begin{equation}
\chi _{H}(z)=\sum\limits_{n=0}^{\infty }D_{n}(z-z_{h})^{n},
\label{horizon chi}
\end{equation}
where the coefficients $D_{n}$ can be calculated order by order from Eq.(\ref{eom}).

To get an analytic solution in the whole regime $[0,z_{h}]$, we smoothly match the two asymptotic solutions $\chi _{H}$ and $\chi _{B}$ at a medium point $z_{\epsilon }\in \left( 0,z_{h}\right)$,
\begin{equation}
\chi _{H}(z_{\epsilon })=\chi _{B}(z_{\epsilon }),~\chi _{H}^{\prime}(z_{\epsilon })=\chi _{B}^{\prime }(z_{\epsilon }).
\end{equation}
For a giving $m_{q}$, the quark condensation can be solved from the above matching equations,
\begin{equation*}
\left. \Sigma =-m_{q}\zeta ^{2}\frac{(\chi _{2}/\chi _{H})^{\prime }}{(\chi_{1}/\chi _{H})^{\prime }}\right\vert _{z=z_{\epsilon }}.
\end{equation*}

The quark condensation vs. temperature at different chemical potentials are plotted in FIG.\ref{SigmaT}(a). For the physical quark mass, quark condensation is not an exact order parameter for $\chi $SB, so that $\Sigma$ smoothly approaches to a small but nonzero value at high temperature. While at low temperature, $\Sigma$ becomes large implying the chiral symmetry breaking. For $\mu < \mu_{c}$, the quark condensation is a monotonous function indicating that $\chi$SB is a crossover. The transition temperature can be determined from the point with the maximum changing rate as shown in FIG.\ref{SigmaT}(b). For $\mu > \mu_{c}$, the quark condensation becomes a multi-valued function, and the phase transition between black holes induces a transition on quark condensation as shown in FIG.\ref{SigmaT}(c). Since we have interpreted the black hole phase transition as the quarkyonic transition in hQCD, this shows that the quarkyonic transition coincides with $\chi$SB and the point $(\mu_{c},T_{c})$ should be identified as the CEP of $\chi$SB.\\

\begin{figure}[tbp]
\subfloat[]{\includegraphics[width=.45\linewidth]{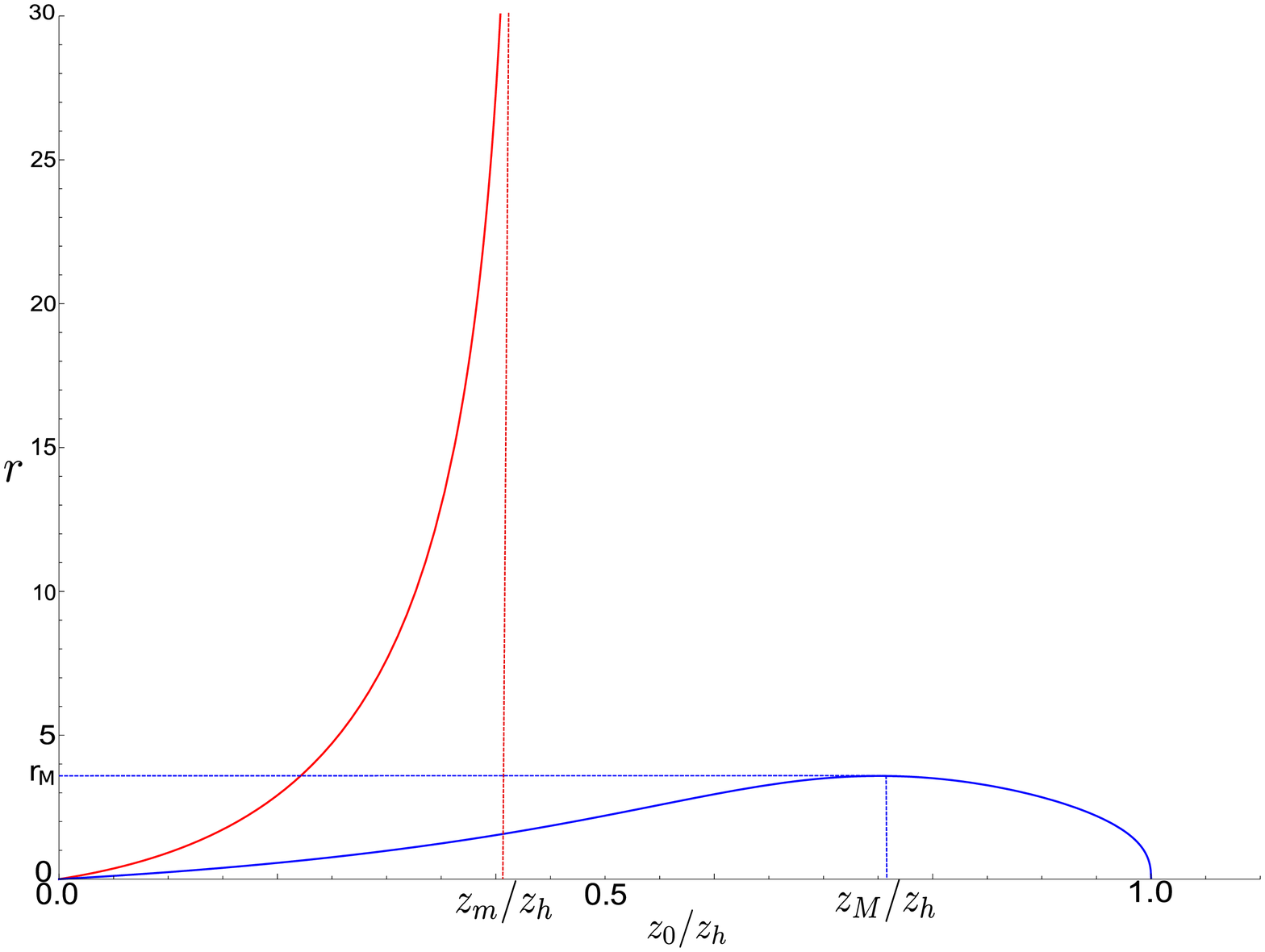}}
\subfloat[]{\includegraphics[width=.48\linewidth]{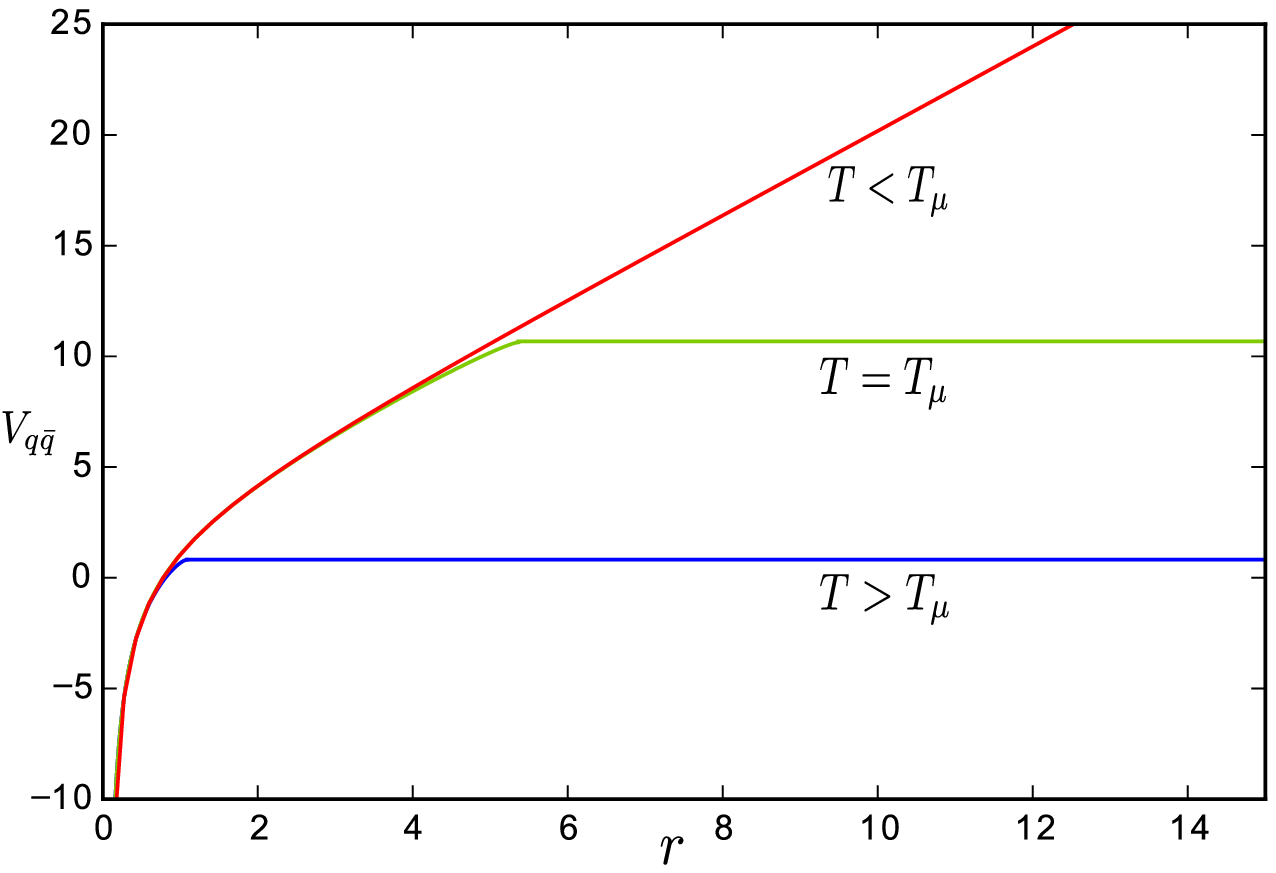}}
\caption{(a) The distant between quark and antiquark $r$ vs. $z_0$. (b) The $q\bar q$ potential at different temperatures.}
\label{Vr}
\end{figure}

\begin{figure}[tbp]
\includegraphics[width=.8\linewidth]{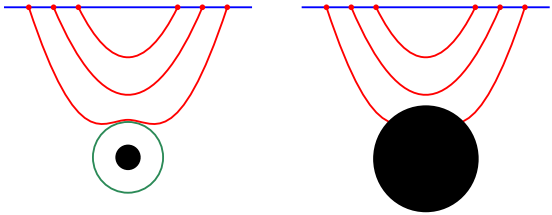}
\caption{The cartoon of the dynamical walls for small and large black holes. The green circle represents the dynamical wall.}
\label{dwall}
\end{figure}

\noindent \textit{Deconfinement Transition} To study the deconfinement transition in hQCD, we add probing open strings in the black hole background and consider the expectation value of the Polyakov loop $\left\langle P\left(\mathcal{C}\right) \right\rangle \sim e^{-V_{q\bar{q}}(r,T)/T}\simeq e^{-S_{on}}$ as the ordered parameter \cite{9803002}, where $V_{q\bar{q}}$ is the quark potential and $S_{on}$ is the on-shell Nambu-Goto action on a world-sheet bounded by a loop $\mathcal{C}$. Choosing the static gauge, $\xi^{0}=t,$ $\xi ^{1}=x$ and the boundary conditions $z\left( \pm r/2\right) =0$, $z(0)=z_{0}$ and $z^{\prime }(0)=0$ with $z_{0}$ the maximum depth that the string can reach, the shape of an open string can be solved from the equation
\begin{equation}
z^{\prime }=\sqrt{g\left( \frac{\sigma ^{2}(z)}{\sigma ^{2}(z_{0})}-1\right)},
\end{equation}
where $\sigma (z)=z^{-2}e^{2A_{s}(z)}\sqrt{g(z)}$ is the effective string tension. The distance and the quark potential are calculated as \cite{1703.09184},
\begin{align}
r& =2\int_{0}^{z_{0}}dz\left[ g(z)\left( \frac{\sigma ^{2}(z)}{\sigma^{2}(z_{0})}-1\right) \right] ^{\frac{1}{2}}, \\
V_{q\bar{q}}& =2\int_{0}^{z_{0}}dz\frac{\sigma (z)}{\sqrt{g(z)}}\left[ 1-\frac{\sigma ^{2}(z_{0})}{\sigma ^{2}(z)}\right] ^{-\frac{1}{2}},
\end{align}
which are plotted in FIG.\ref{Vr}.
At low temperature or small black hole (red line), $r\to \infty$ at $z_0=z_m$ where the dynamical wall emerges. The quark potential is linear at large $r$ and open strings are always connected to form bounded states, so that QCD is in the confinement phase. While at high temperature or large black hole (blue line), there exists a maximum distance $r=r_M$ beyond that open strings break and the quark potential becomes constant, so that quarks and antiquarks become free indicating that QCD is in the deconfinement phase. A cartoon of the dynamical wall for small and large black holes is shown in FIG.\ref{dwall}. The dynamical wall is reminiscence of the IR cut-off in the hard wall model, but here the wall emerges dynamically instead of imposing by hand.

\begin{figure}[tbp]
\subfloat[]{\includegraphics[width=.45\linewidth]{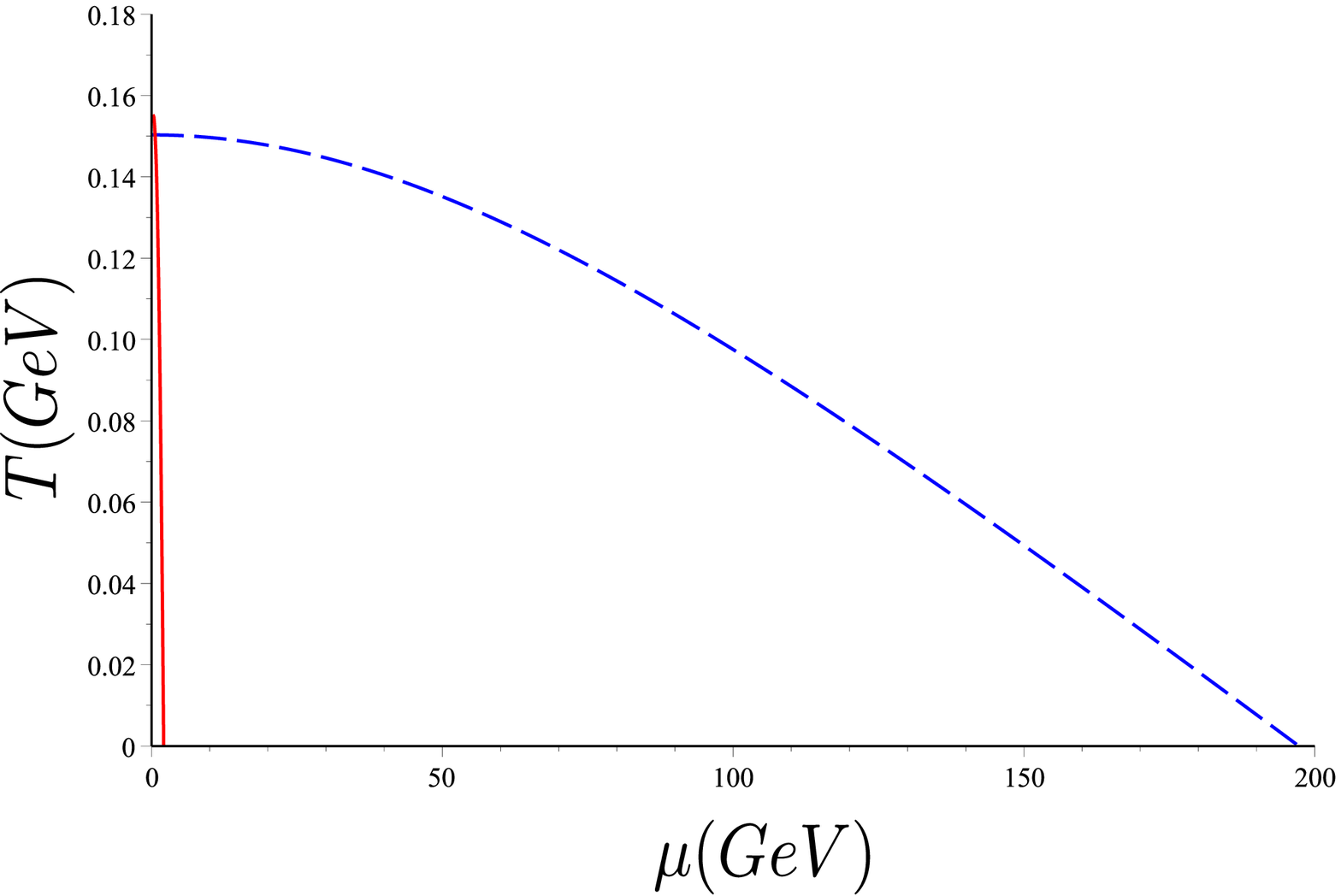}} 
\subfloat[]{\includegraphics[width=.45\linewidth]{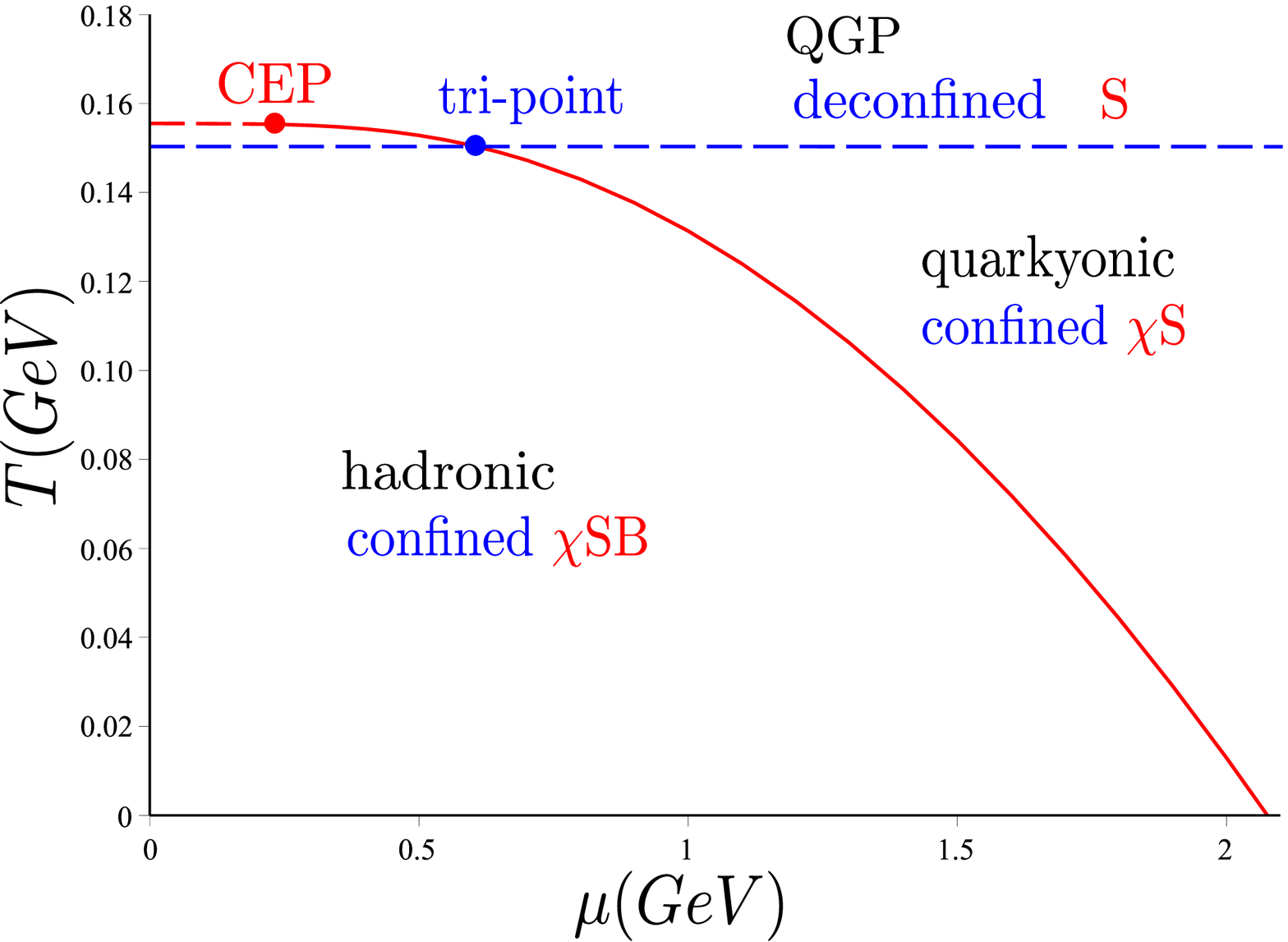}}
\caption{(a) The deconfinement (dashed blue line) and $\protect\chi$SB (solid red line) transitions. We enlarge the region near $\protect\mu\sim0$ in (b).}
\label{Tmustring}
\end{figure}

The dynamical wall is a crucial concept to understand the deconfinement transition in the holographic framework. It disappears at the deconfinement transition temperature. The phase diagram of the deconfinement transition (dashed blue line) is plotted in FIG.\ref{Tmustring}(a). Since the dynamical wall appears as the black hole horizon decreasing and passing the critical horizon smoothly, we deduce that the deconfinement transition is a crossover. In addition, We notice that the deconfinement transition temperature depends on the chemical potential, but the dependence is much weaker than the $\chi$SB. To comparing, we also plot the $\chi$SB line (red curve) in FIG.\ref{Tmustring}(a). The deconfinement and $\chi$SB lines intersect at a tri-point $(\mu_t, T_t)$. The near $\mu=0$ region of the QCD phase diagram is enlarged and plotted in FIG.\ref{Tmustring}(b).\\

\noindent \textit{QCD Phase Diagram} The phase diagram FIG.\ref{Tmustring}(b) is consistent with the conjectured QCD phase diagram at finite $N_{c}$ as shown in FIG.\ref{Nphase}(b). The deconfinement line is a crossover and slightly bends down with $\mu$. The quarkyonic line coincides with the $\chi$SB line. There exists the quarkyonic phase between the deconfinement and $\chi$SB lines. The QGP, hadronic and quarkyonic phases encounter at a tri-point which is close to the chiral CEP.

\begin{figure}[tbp]
\subfloat[$\mu<\mu_t$]{\includegraphics[width=.45\linewidth]{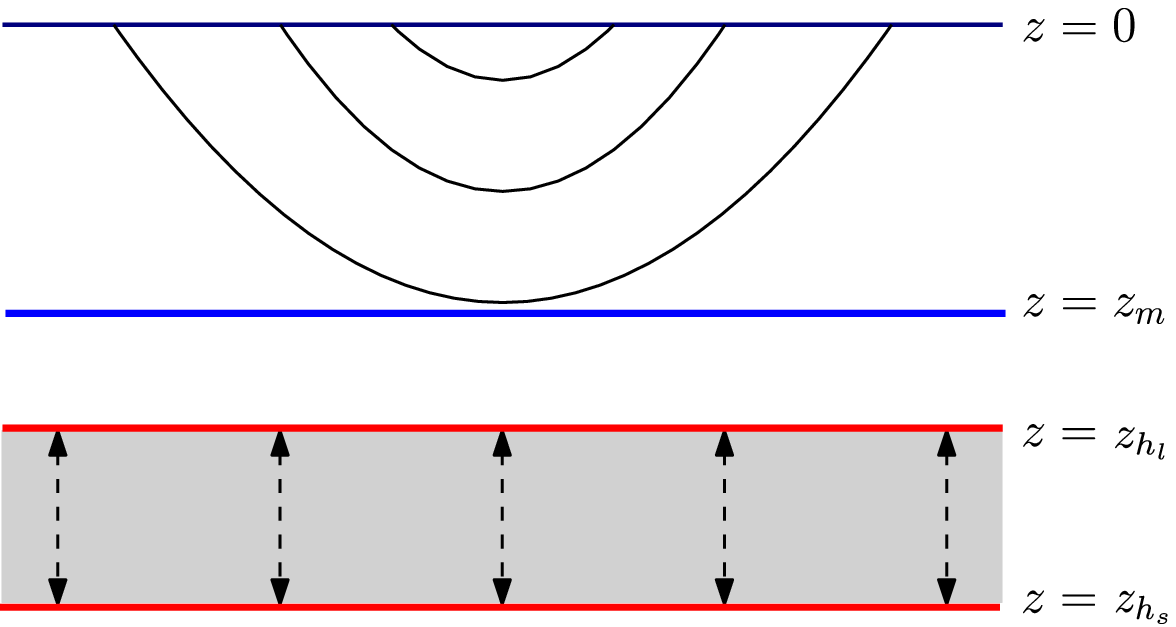}
\includegraphics[width=.45\linewidth,height=.25\linewidth]{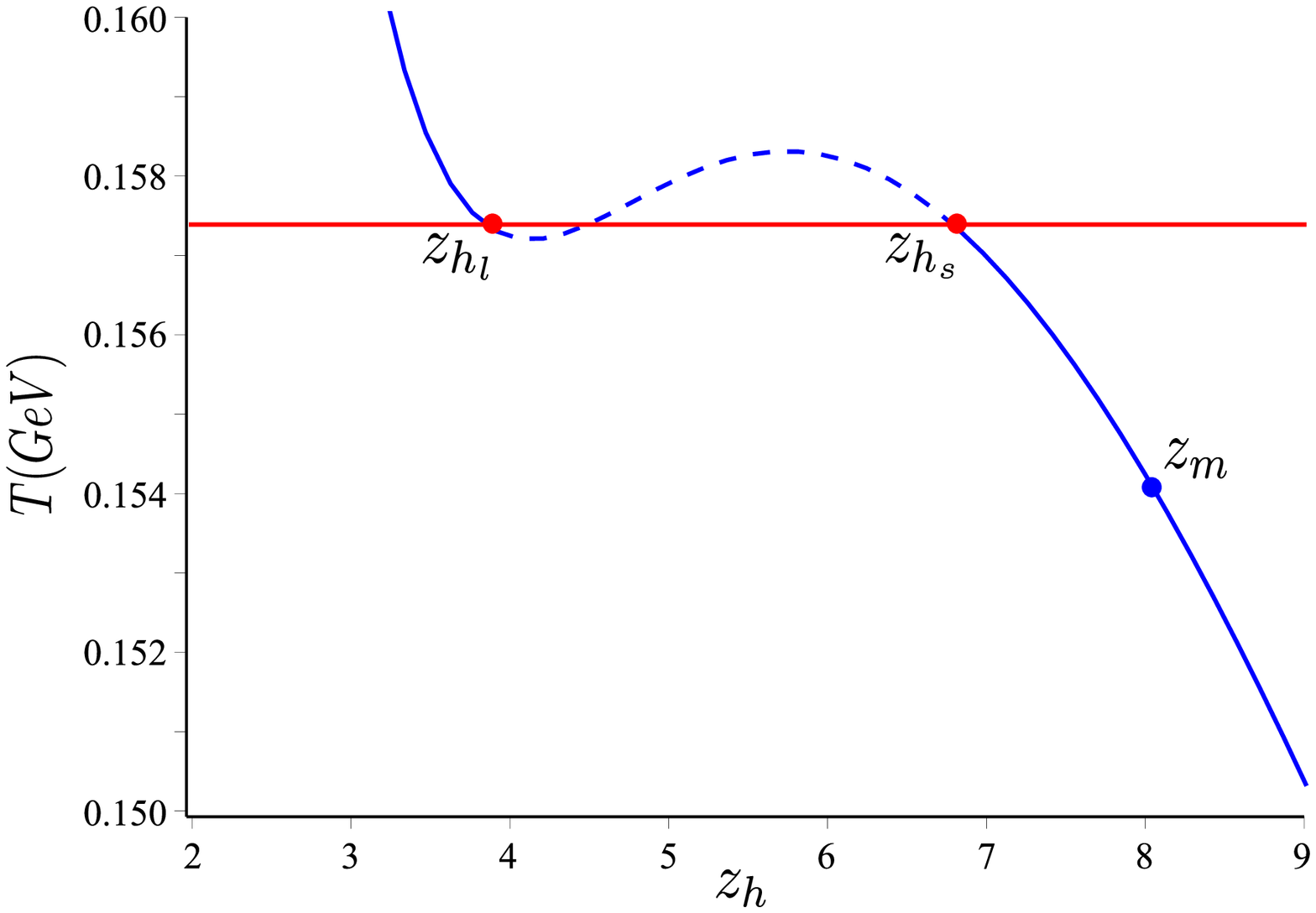}}\\
\subfloat[$\mu>\mu_t$]{\includegraphics[width=.45\linewidth]{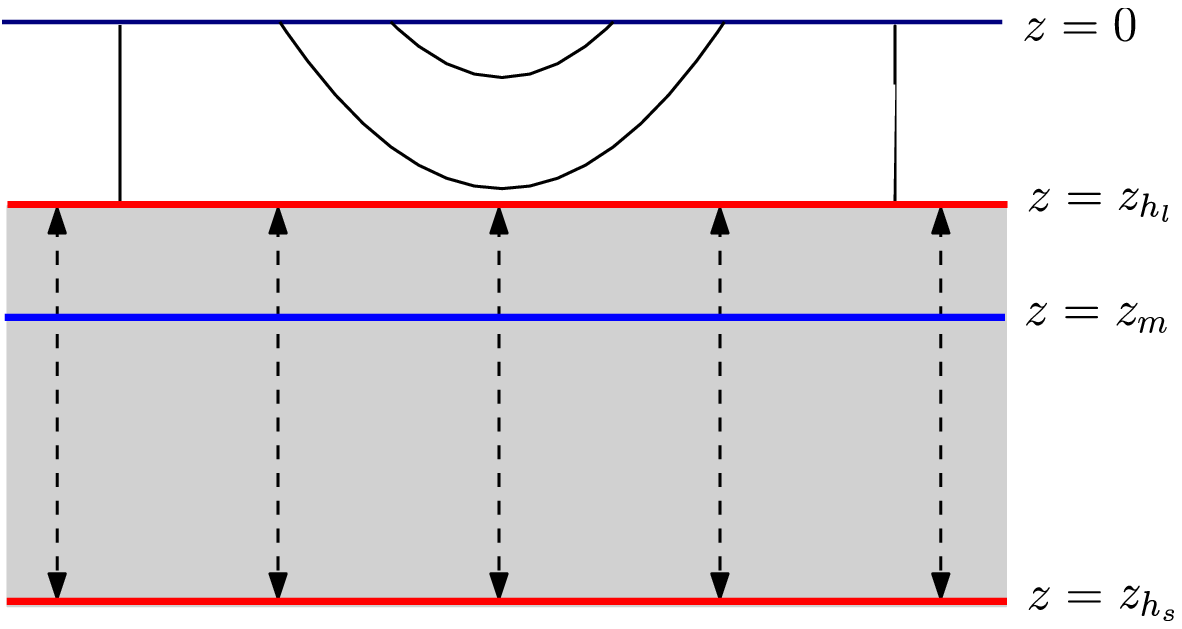}
\includegraphics[width=.45\linewidth,height=.25\linewidth]{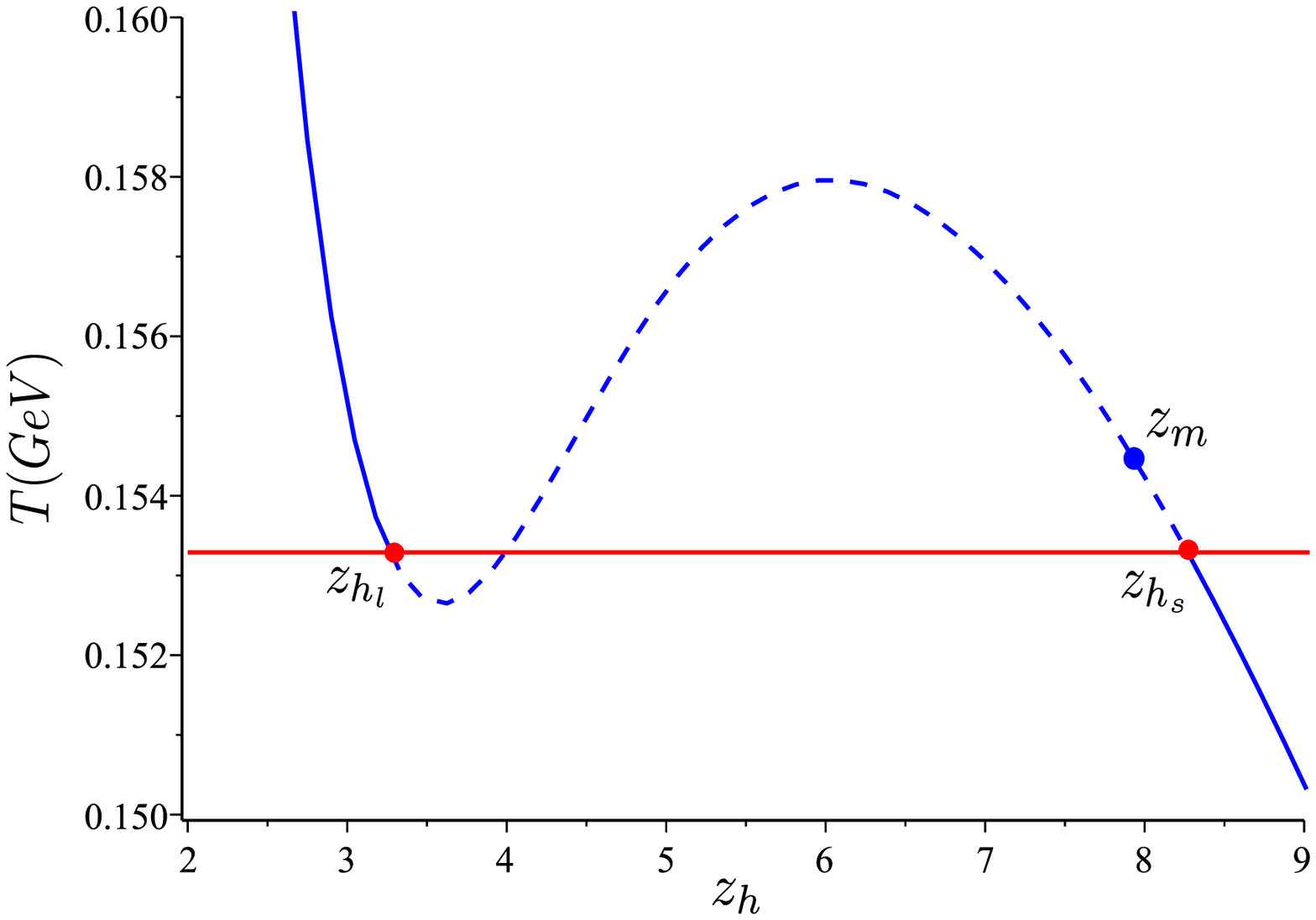}}
\caption{The cartoon for jumping the dynamical wall and the corresponding temperature at small and large chemical potentials.}
\label{dwallBH}
\end{figure}

\begin{figure}[tbp]
\subfloat[]{\includegraphics[width=.45\linewidth]{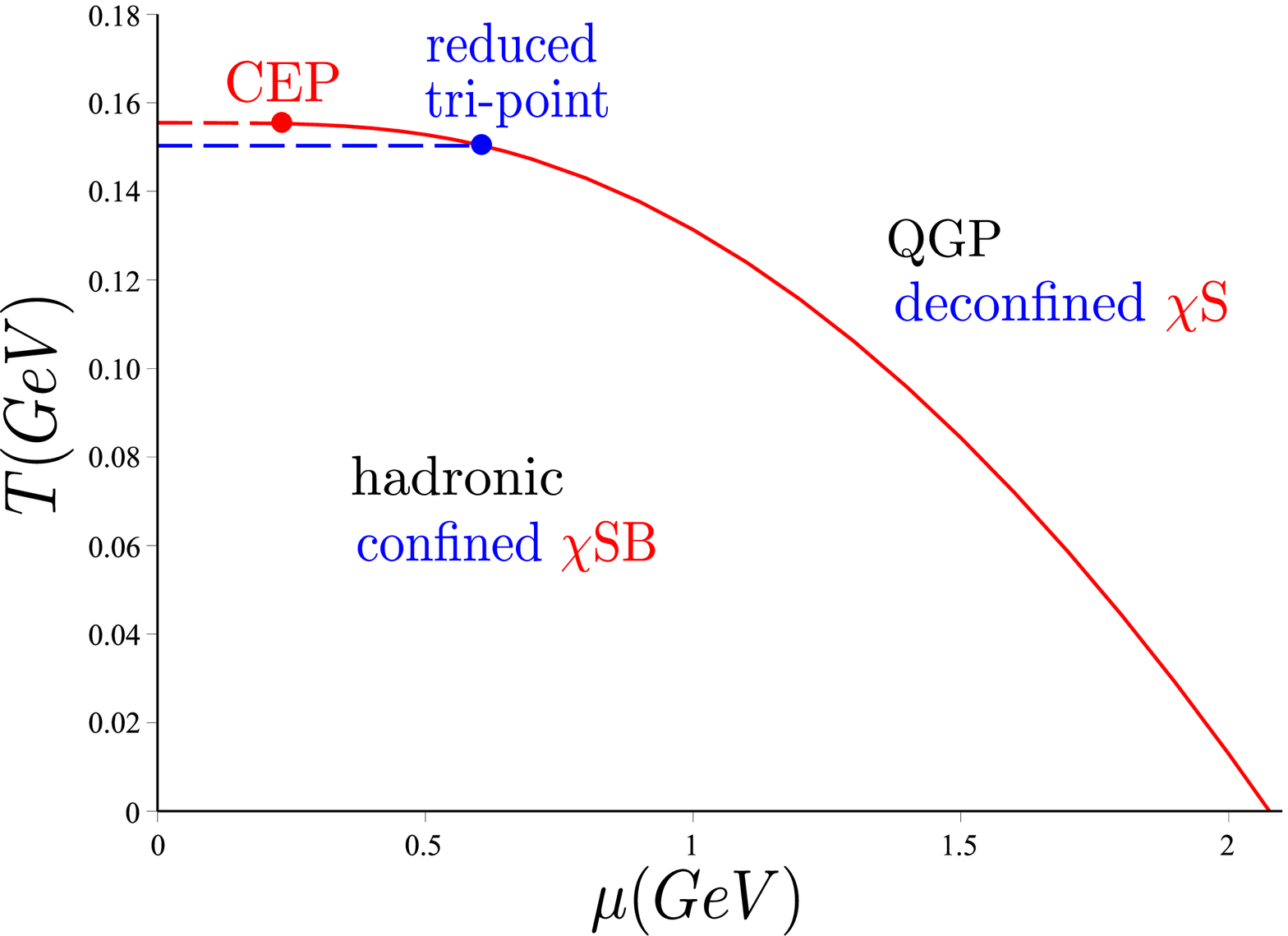}} 
\subfloat[]{\includegraphics[width=.45\linewidth]{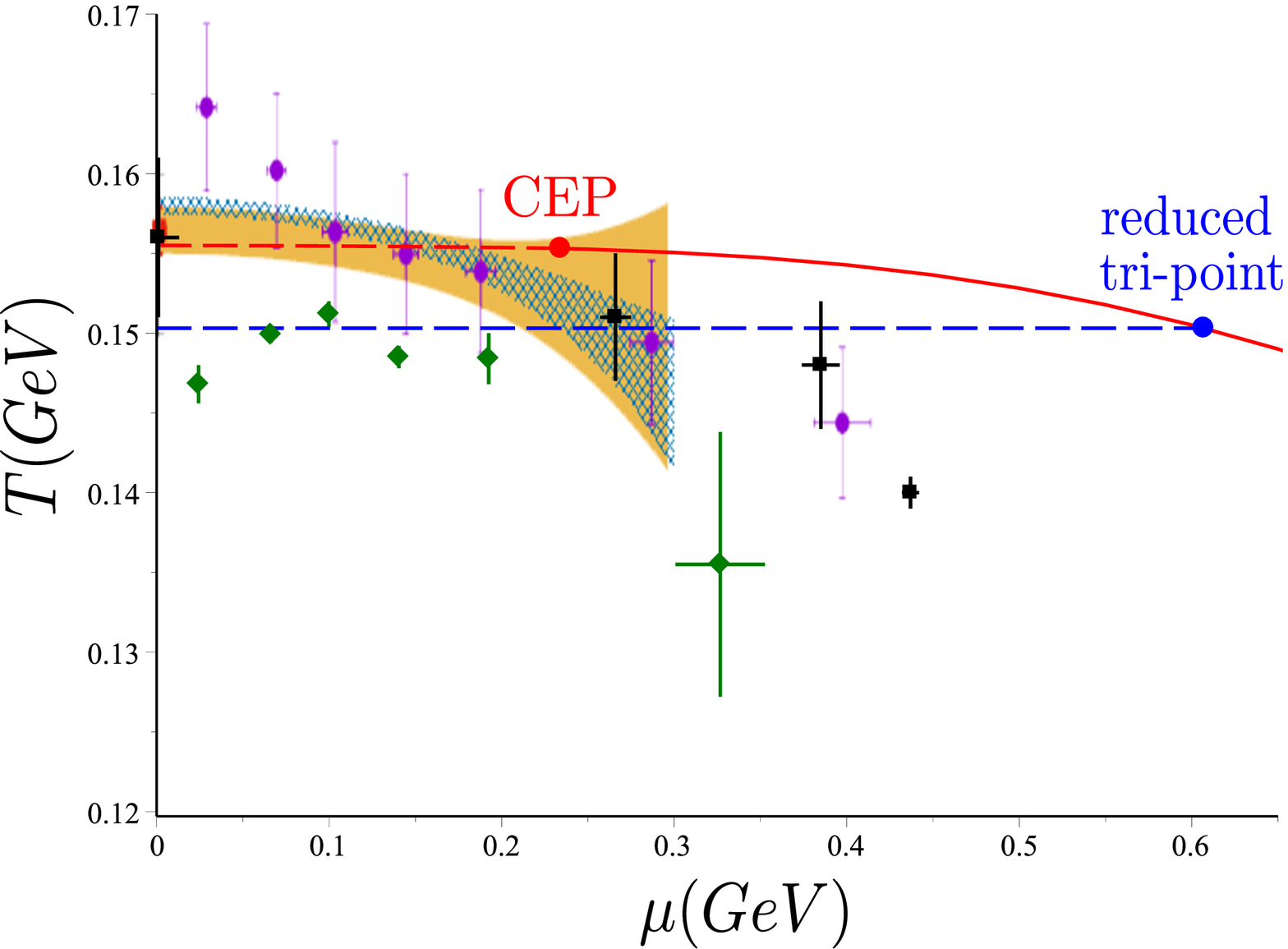}}
\caption{(a) The conclusive QCD phase diagram. The blue and red lines represent the deconfinement and $\chi$SB transitions, which intersect at the reduced tri-point at $(\mu_t,T_t)\simeq(0.607, 0.150)$. The transition temperatures for deconfinement and $\chi$SB at $\mu=0$ are $T_d\simeq 0.1503$ and $T_\chi \simeq 0.1555$. The $\chi$SB CEP is at $(\mu_c,T_c)\simeq(0.234, 0.155)$. (b) The chemical freeze-out parameters from the STAR and ALICE collaborations are plotted in black squares \cite{1212.2431}, green diamonds \cite{1403.4903}, purple circles \cite{1701.07065} and the red square \cite{1710.09425}. The chiral crossover temperature obtained  up to the order of $O\left(\mu^4/T^4\right)$ from the HotQCD \cite{1812.08235,2002.11957} and Wuppertal-Budapest collaborations \cite{2002.02821} are marked with the yellow and blue-net zones.} 
\label{final}
\end{figure}

Nevertheless, this is not the final result for the QCD phase diagram yet. In fact, the black hole phase transition dramatically affects the deconfinement transition in an novel way. To explain it, we plot the jumps of the black hole horizon in FIG.\ref{dwallBH}. For $\mu <\mu_{t}$, the jump of the horizon between $z=z_{h_{s}}$ and $z=z_{h_{l}}$ (red lines or dots) during the black hole phase transition does not affect the dynamical wall (blue line or dot) at $z=z_{m}$. While for $\mu >\mu_{t}$, the jump of the horizon bypasses the dynamical wall and forces the deconfinement transition to take place at the same temperature of the black hole transition. Therefore, the deconfinement line coincides with the quarkyonic phase line (thereby also coincides with the $\chi$SB line) and the quarkyonic phase disappears. More exactly, the deconfinement transition occurs in the dynamically unstable region which is bypassed by the black hole phase transition, as shown in FIG.\ref{dwallBH}(b).

The final QCD phase diagram is plotted in FIG.\ref{final}(a). At $\mu = 0$, both deconfinement and $\chi$SB are crossovers with $T_{d}\lesssim T_{\chi}$. The $\chi$SB becomes first order at the CEP, and the deconfinement becomes first order at the reduced tri-point beyond that the deconfinement transition coincides with $\chi$SB. The chemical freeze out (CFO) parameters extracted from the experimental measurements in heavy-ion collisions are plotted in FIG.\ref{final}(b) to compare with the phase diagram obtained from our holographic QCD model. The CFO is due to the rapid decrease of degree of freedom in the system. At $T=0$, $\mu_{CFO}\simeq 0.938$ represents nuclear mater in the ground state so that the CFO temperature should be less than the $\chi$SB temperature for large $\mu$ \cite{1105.3934}. While in the crossover region $\mu<\mu_c$, it was showed that the CFO temperature is close to the $\chi$SB temperature \cite{1909.02991,2002.11957}. As shown in  FIG.\ref{final}(b), the chemical freeze-out parameters from the STAR and ALICE collaborations are consistent with the phase diagram obtained from our model.\\

\noindent \textit{Conclusion} In this work, we investigated QCD phase diagram in a well studied hQCD model constructed in the 5-dimensional Einstein-Maxwell-scalar system. To realize the quarkyonic phase in the holographic correspondence, we proposed to interpret the black hole phase transition in the bulk spacetime as the quarkyonic transition in the dual QCD theory and showed that it coincides with the $\chi$SB. We justified our proposal by showing the significantly jump of the baryon density, which is conjectured to be the order parameter for the quarkyonic transition, at the black hole transition temperature. In addition, we introduced the bypass mechanism to explain why the deconfinement line coincides with the $\chi$SB line despite that their physical origins are quite different. The final QCD phase diagram is shown in FIG.\ref{final}(a). The chemical freeze-out parameters from the STAR and ALICE collaborations are well consistent with the phase diagram we obtained from our holographic QCD model as shown in  FIG.\ref{final}(b).

\section*{Acknowledgements}

We would like to thank Song He, Mei Huang, Danning Li, Wen-Yu Wen for useful discussions. This work of YY is supported by the Ministry of Science and Technology (MOST 106-2112-M-009 -005 -MY3) and National Center for Theoretical Science, Taiwan. The work of PHY was supported by the University of Chinese Academy of Sciences.


\begin{thebibliography}{99}
\bibitem {Cabibbo} Cabibbo N and Parisi G 1975 Phys. Lett. B 59 67.

\bibitem{0609068} Y. Aoki, Z. Fodor, S.D. Katz, K.K. Szabo, "The QCD transition temperature: results with physical masses in the continuum limit", arXiv:hep-lat/0609068, Phys.Lett.B643:46-54,2006.

\bibitem {1111.4953}Michael Fromm, Jens Langelage, Stefano Lottini, 
Owe Philipsen, "The QCD deconfinement transition for heavy quarks and all baryon chemical potentials", arXiv:1111.4953 [hep-lat], JHEP 01 (2012) 042.

\bibitem{0706.2191} Larry McLerran, Robert D. Pisarski, "Phases of 
Dense Quarks at Large $N_{c}$", arXiv:0706.2191, Nucl.Phys.A796:83-100,2007.

\bibitem{1801.09215} Zhibin Li, Kun Xu, Xinyang Wang, Mei Huang, 
"The kurtosis of net baryon number fluctuations from a realistic Polyakov--Nambu--Jona-Lasinio model along the experimental freeze-out line", arXiv:1801.09215 [hep-ph], Eur.Phys.J.C 79 (2019) 3, 245.

\bibitem{0701022} Troels Harmark, Vasilis Niarchos, Niels A. Obers, 
"Instabilities of Black Strings and Branes", arXiv:hep-th/0701022, Class.Quant.Grav.24:R1-R90,2007.

\bibitem{0803.3318} Kenji Fukushima, 
"Phase diagrams in the three-flavor Nambu--Jona-Lasinio model with the Polyakov loop", arXiv:0803.3318 [hep-ph], Phys.Rev.D77:114028,2008.

\bibitem{0805.1509} H. Abuki, R. Anglani, R. Gatto, G. Nardulli, M. Ruggieri, 
"Chiral crossover, deconfinement and quarkyonic matter within a Nambu-Jona Lasinio model with the Polyakov loop", arXiv:0805.1509 [hep-ph], Phys.Rev.D78:034034,2008.

\bibitem{0911.4806} A. Andronic, D. Blaschke, P. Braun-Munzinger, J. Cleymans, K. Fukushima, L.D. McLerran, H. Oeschler, R.D. Pisarski, K. Redlich, C. Sasaki, H. Satz, J. Stachel, 
"Hadron Production in Ultra-relativistic Nuclear Collisions: Quarkyonic Matter and a Triple Point in the Phase Diagram of QCD", arXiv:0911.4806 [hep-ph], Nucl.Phys.A837:65-86,2010.

\bibitem{0905.2949} C.E. DeTar, U.M. Heller, 
"QCD Thermodynamics from the Lattice", arXiv:0905.2949 [hep-lat], Eur.Phys.J.A 41 (2009) 405-437. 

\bibitem{1104.0873} Fukun Xu, Hong Mao, Tamal K. Mukherjee, Mei Huang, 
"Dressed Polyakov loop and flavor dependent phase transitions", arXiv:1104.0873 [hep-ph], Phys.Rev.D 84 (2011) 074009. 

\bibitem{1602.06699} M. Shifman, 
"Dynamically Emergent Flavor in a Confining Theory with Unbroken Chiral Symmetry", arXiv:1602.06699 [hep-ph], Phys. Rev. D 93, 074035 (2016).

\bibitem{1705.00718} Kenji Fukushima, Vladimir Skokov, 
"Polyakov loop modeling for hot QCD", arXiv:1705.00718 [hep-ph], Prog.Part.Nucl.Phys. 96 (2017) 154-199.

\bibitem{1405.1289} Takahiro M. Doi, Hideo Suganuma, Takumi Iritani, 
"Relation between Confinement and Chiral Symmetry Breaking in Temporally Odd-number Lattice QCD", arXiv:1405.1289 [hep-lat], Phys. Rev. D 90, 094505 (2014).

\bibitem{1502.07706} M. Pak, M. Schröck, 
"Overlap Quark Propagator in Coulomb Gauge QCD and the Interrelation of Confinement and Chiral Symmetry Breaking", arXiv:1502.07706 [hep-lat], Phys. Rev. D 91 (2015) 074515.

\bibitem{1709.05981} Hideo Suganuma, Takahiro M. Doi, Krzysztof Redlich, Chihiro Sasaki, 
"Relating Quark Confinement and Chiral Symmetry Breaking in QCD", arXiv:1709.05981 [hep-lat], J.Phys. G44 (2017) no.12, 124001. 

\bibitem {9711200} Juan M. Maldacena, 
"The Large N limit of superconformal field theories and supergravity", Int.J.Theor.Phys. 38 (1999) 1113-1133, hep-th/9711200.

\bibitem{1511.02721} Kaddour Chelabi, Zhen Fang, Mei Huang, Danning Li, Yue-Liang Wu, 
"Realization of chiral symmetry breaking and restoration in holographic QCD", arXiv:1511.02721 [hep-ph], Phys. Rev. D 93, 101901 (2016). 

\bibitem{1512.06493}, Kaddour Chelabi, Zhen Fang, Mei Huang, Danning Li, Yue-Liang Wu, 
"Chiral Phase Transition in the Soft-Wall Model of AdS/QCD", arXiv:1512.06493 [hep-ph], JHEP 04 (2016) 036.

\bibitem{1610.09814} Danning Li, Mei Huang, 
"Chiral phase transition of QCD with Nf=2+1 flavors from holography", arXiv:1610.09814 [hep-ph], JHEP 02 (2017) 042.

\bibitem{1810.07019} Jianwei Chen, Song He, Mei Huang, Danning Li, 
"Critical exponents of finite temperature chiral phase transition in soft-wall AdS/QCD models", arXiv:1810.07019 [hep-ph], JHEP 01 (2019) 165. 

\bibitem{1908.02000} Xun Chen, Danning Li, Defu Hou, Mei Huang, 
"Quarkyonic phase from quenched dynamical holographic QCD model", arXiv:1908.02000 [hep-ph], JHEP 03 (2020) 073. 

\bibitem{1910.02383} Yoon-Seok Choun, Sang-Jin Sin, 
"Bridging the Chiral symmetry and Confinement with Singularity", arXiv:1910.02383 [hep-th], Phys.Lett.B 805 (2020) 135433.

\bibitem{1301.0385} Song He, Shang-Yu Wu, Yi Yang, Pei-Hung Yuan, 
"Phase Structure in a Dynamical Soft-Wall Holographic QCD Model", arXiv:1301.0385 [hep-th], JHEP 04 (2013) 093.

\bibitem{1406.1865} Yi Yang, Pei-Hung Yuan, 
"A Refined Holographic QCD Model and QCD Phase Structure", arXiv:1406.1865 [hep-th], JHEP 11 (2014) 149.

\bibitem{1506.05930} Yi Yang, Pei-Hung Yuan, 
"Confinement-Deconfinment Phase Transition for Heavy Quarks", arXiv:1506.05930 [hep-th], JHEP 12 (2015) 161.

\bibitem{1703.09184} Meng-Wei Li, Yi Yang, Pei-Hung Yuan, 
"Approaching Confinement Structure for Light Quarks in a Holographic Soft Wall QCD Model", arXiv:1703.09184 [hep-th], Phys.Rev.D 96 (2017) 6, 066013 .

\bibitem{1705.07587} Yi Yang, Pei-Hung Yuan, 
"Universal Behaviors of Speed of Sound from Holography", arXiv:1705.07587 [hep-th], Phys.Rev.D 97 (2018) 12, 126009.

\bibitem{2004.01965} Song He, Yi Yang, Pei-Hung Yuan, 
"Analytic Study of Magnetic Catalysis in Holographic QCD", arXiv:2004.01965 [hep-th].

\bibitem{2009.05694} Meng-Wei Li, Yi Yang, Pei-Hung Yuan, 
"Analytic Study on Chiral Phase Transition in Holographic QCD", arXiv:2009.05694 [hep-th].

\bibitem{2002.02821} Szabolcs Borsanyi, Zoltan Fodor, Jana N. Guenther, Ruben Kara, Sandor D. Katz, Paolo Parotto, Attila Pasztor, Claudia Ratti, Kalman K. Szabo, 
"The QCD crossover at finite chemical potential from lattice simulations", arXiv:2002.02821 [hep-lat], Phys. Rev. Lett. 125, 052001 (2020).

\bibitem{1812.09676} Meng-Wei Li, Yi Yang, Pei-Hung Yuan, 
"Imprints of Early Universe on Gravitational Waves from First-Order Phase Transition in QCD", arXiv:1812.09676 [hep-th].

\bibitem{9803002} Juan M. Maldacena, 
"Wilson loops in large N field theories", arXiv:hep-th/9803002, Phys.Rev.Lett. 80 (1998) 4859-4862.

\bibitem{1212.2431} Francesco Becattini, Marcus Bleicher, Thorsten Kollegger, Tim Schuster, Jan Steinheimer, Reinhard Stock, 
"Hadron Formation in Relativistic Nuclear Collisions and the QCD Phase Diagram", arXiv:1212.2431 [nucl-th], Phys. Rev. Lett. 111 (2013) 082302.

\bibitem{1403.4903} Paolo Alba, Wanda Alberico, Rene Bellwied, Marcus Bluhm, Valentina Mantovani Sarti, Marlene Nahrgang, Claudia Ratti, 
"Freeze-out conditions from net-proton and net-charge fluctuations at RHIC", arXiv:1403.4903 [hep-ph], Phys. Lett. B 738 (2014) 305-310.

\bibitem{1701.07065} STAR Collaboration, 
"Bulk Properties of the Medium Produced in Relativistic Heavy-Ion Collisions from the Beam Energy Scan Program", arXiv:1701.07065 [nucl-ex], Phys. Rev. C 96 (2017) 4, 044904.

\bibitem{1710.09425} A. Andronic, P. Braun-Munzinger, K. Redlich, J. Stachel, 
"Decoding the phase structure of QCD via particle production at high energy", arXiv:1710.09425 [nucl-th], Nature 561 (2018) 7723, 321-330.

\bibitem{1812.08235} A. Bazavov, H.-T. Ding, P. Hegde, O. Kaczmarek, F. Karsch, N. Karthik, E. Laermann, Anirban Lahiri, R. Larsen, S.-T. Li, Swagato Mukherjee, H. Ohno, P. Petreczky, H. Sandmeyer, C. Schmidt, S. Sharma, P. Steinbrecher, 
"Chiral crossover in QCD at zero and non-zero chemical potentials", arXiv:1812.08235 [hep-lat], Phys.Lett.B 795 (2019) 15-21.

\bibitem{2002.11957} Heng-Tong Ding, 
"New developments in lattice QCD on equilibrium physics and phase diagram", arXiv:2002.11957 [hep-lat], Quark Matter 2019.

\bibitem{1105.3934} Sourendu Gupta, Xiaofeng Luo, Bedangadas Mohanty, Hans Georg Ritter, Nu Xu, 
"Scale for the Phase Diagram of Quantum Chromodynamics", arXiv:1105.3934 [hep-ph], Science 332 (2011) 1525-1528.

\bibitem{1909.02991} Wei-jie Fu, Jan M. Pawlowski, Fabian Rennecke, 
"The QCD phase structure at finite temperature and density", arXiv:1909.02991 [hep-ph], Phys. Rev. D 101, 054032 (2020).



\end{thebibliography}
\end{document}